\documentclass{article}
\usepackage[spanish,english,es-tabla]{babel}
\usepackage[utf8]{inputenc}
\usepackage[T1]{fontenc}
\usepackage{lmodern}     
\usepackage[protrusion=true, expansion=true]{microtype}  
\usepackage{amsmath,amssymb} 
\usepackage{tabularx}    
\usepackage{multicol}    
\usepackage{multirow}    
\usepackage{longtable}   
\usepackage{booktabs}    
\usepackage{ltablex}
\usepackage{setspace} 
\usepackage{tikz}                
\usepackage{graphicx,graphics}   
\usepackage{subfigure}           
\usepackage[export]{adjustbox}
\usepackage{ragged2e} 
\usepackage{blindtext}
\usepackage[acronym,nonumberlist,nomain]{glossaries}  
\usepackage{hyperref}
\usepackage[a4paper]{geometry}
\usepackage{parskip}     
\usepackage{fancyhdr}    
\usepackage{enumitem}
\usepackage{xcolor}
\usepackage{emptypage}   
\usepackage[small,hang,bf]{caption}    
\usepackage{subcaption}   
\usepackage[ 
    backend=biber,
    natbib=true,
    style=numeric,
    sorting=none,
    maxnames=99,
    maxcitenames=99
]{biblatex}
\usepackage{csquotes}
\usetikzlibrary{arrows.meta}
\usepackage{algpseudocode}
\usepackage{pgfplots}
\usepackage{wasysym}
\pgfplotsset{compat=1.18}
\usepackage{changes}
\usepackage[noblocks]{authblk}
\usepackage{wasysym}
\DeclareGraphicsExtensions{.pdf,.png,.jpg}

\usepackage{cancel}

\usepackage{xstring} 

\title{A Correction Method for Crack Area Overestimation in Phase-Field Fracture}
\author[1,2]{M. Castillón}
\author[3,2]{J. Segurado}
\author[1,2]{I. Romero}
\affil[1]{Dept. of Mechanical Engineering, Universidad Politécnica de Madrid, Spain}
\affil[2]{IMDEA Materials Institute, Getafe, Madrid, Spain}
\affil[3]{Dept. of Materials Sciences, Universidad Politécnica de Madrid, Spain}

\begin{document}
\maketitle
\begin{abstract}
Phase-field fracture models are known to overestimate the crack area, a discrepancy that compromises the accuracy of fracture predictions. This issue stems from the diffuse crack representation and numerical artifacts, such as strain localization, where the phase-field variable artificially saturates across finite elements.
Existing correction strategies, including mesh-dependent factors and skeletonization algorithms, have significant limitations. Mesh-based corrections are often unreliable for unstructured meshes, while skeletonization can be complex and inaccurate for intricate crack topologies, especially in three dimensions.
This paper introduces a novel and robust framework to correct this overestimation. Our approach is founded on the principle of energy equipartition, where the energy contributions from the phase-field and its gradient are equal as the length-scale parameter approaches zero. Since numerical artifacts primarily affect the phase-field term while leaving the gradient term largely unperturbed, we propose that the crack area can be accurately approximated as twice the gradient-dependent energy. This method is inherently mesh-independent and readily applicable to the entire domain, including 3D simulations.
The proposed methodology is validated against benchmarks with analytical solutions and compared with established methods like skeletonization to demonstrate its accuracy. It is then applied to complex geometries with curvilinear crack paths and evaluated in a three-dimensional simulation.
\end{abstract}

\section{Introduction}
\label{sec:Introduction}

Phase-field fracture models provide a powerful computational framework for simulating complex crack
patterns without the need for explicitly tracking the crack surfaces. These models, which regularize
sharp crack discontinuities into diffuse interfaces, are built on the foundational work of Francfort
and Marigo~\cite{phase_field_FrancfortMarigo1998}, who first proposed a variational formulation of
brittle fracture. This framework was subsequently developed into the modern phase-field method by
researchers such as Bourdin et al.~\cite{phase_field_Bourdin2000} and Miehe et
al.~\cite{phase_field_Miehe2010}. By representing damage with a continuous scalar field, denoted as
$\phi$ in this work, the method naturally captures complex phenomena like crack initiation, propagation, branching, and merging, making it a versatile tool for fracture analysis.

In phase-field fracture modeling, the regularization is governed by a length-scale parameter which
we indicate as $l$ that controls the width of the diffuse crack representation. A central challenge in the finite element implementation of these models lies in the conflicting requirements placed on $l$ and the characteristic mesh size, $h$.
On one hand, the mesh must be fine enough to resolve the phase-field profile. A common guideline,
particularly for the widely used AT2 model \cite{phase_field_Miehe_lh_relation}, is to maintain a ratio of $l/h \ge 2$.
This ensures that the numerical solution can accurately capture the
diffuse crack interface. On the other hand, to approximate the sharp-crack limit of fracture
mechanics, $l$ must be as small as possible. However, this requirement is complicated by a numerical
artifact known as \emph{strain localization} \cite{phase_field_FrancfortMarigo1998, phase_field_effective_Gc_factor_2}. This
phenomenon causes the phase-field variable to artificially saturate at $\phi=1$ across entire finite
elements, distorting the ideal diffuse profile and leading to a significant overestimation of the
crack surface area. As demonstrated in our previous work~\cite{phase_field_Castillon2025}, this overestimation systematically compromises the accuracy of simulation outputs, including the forces, and displacements.

Several strategies have been developed to mitigate this overestimation. Bourdin et al.
\cite{phase_field_Bourdin2008} proposed a correction factor that scales the effective energy
release rate as a function of the ratio $h/l$, leading to $G_c^{\text{eff}} = G_c (1 + h/2l)$. While
simple, its effectiveness relies on the ratio $h/l$ approaching zero, which demands computationally
expensive mesh refinement. Furthermore, this correction is derived under an idealized one-dimensional assumption, and relies on a fixed element size around the crack tip. This assumption can be unreliable for unstructured meshes or complex crack paths. 

It must be noted that the strain localization effect is typically taken into account by modifying the critical energy release rate. However, as presented in \cite{phase_field_Castillon2025}, it can also be handled by addressing the overestimation of the crack area. In this work, this effect is taken into account to correct the crack area directly. We also review these alternative formulations, explaining that all are correct if the equations are treated in a consistent manner.

Alternative post-processing techniques, such as skeletonization algorithms \cite{phase_field_skeleton, phase_field_Castillon2025}, measure
the crack area by applying a threshold to the computed phase-field (e.g., identifying regions where
$\phi > 0.9$). Although effective for 2D simulations, their accuracy is sensitive to the chosen
threshold and mesh quality. Moreover, these methods are computationally intensive and their
extension to 3D is complex. 
As a result of these considerations, two competing requirements arise.  To approach the sharp-crack limit, the length-scale parameter $l$ must be small. However, the mesh size $h$ must be fine enough to resolve the phase-field profile, while also being small relative to $l$ to minimize the strain localization error, which is proportional to $h/l$. This often leads to computationally prohibitive mesh densities, highlighting the need for more robust and efficient correction methods.

Another significant artifact in phase-field fracture simulations is the presence of a non-physical peak force at the onset of crack initiation, a phenomenon not observed in experimental results~\cite{phase_field_snap_pedro, phase_field_snap_Ritukesh, phase_field_snap_Zambrano}. This force overshoot is a numerical artifact linked to the nucleation process, where the phase-field variable evolves from an undamaged state ($\phi=0$) to a fully damaged state ($\phi=1$). This issue is addressed by existing correction methods with varying degrees of success. The Bourdin correction, being a constant scaling factor dependent on mesh size and length scale, reduces the magnitude of the peak force but does not eliminate the overshoot itself. In contrast, skeletonization-based methods, which measure the crack length by thresholding the phase-field (e.g., considering only regions where $\phi > 0.95$), effectively bypass the nucleation phase and thus remove the peak force from the corrected results. However, this approach has its own limitations: its accuracy depends on the chosen threshold and image processing techniques, and its application to complex crack topologies, such as multiple interacting cracks or three-dimensional problems, is not straightforward.

This paper introduces a novel correction method to address the overestimation of crack area
calculation in phase-field fracture simulations. Our approach is founded on the equipartition of the fracture-energy components. According to this result, the energy contributions from the phase-field variable and its
gradient are equal when the diffuse crack is well developed and free from boundary effects. We leverage this property to counteract the effects of strain localization. This numerical artifact artificially inflates the phase-field energy by forcing $\phi=1$ across entire elements, while the gradient energy term remains largely unperturbed since $\nabla\phi=0$ in these saturated regions.

Given that the gradient contribution to the fracture energy is less sensitive to localization, we
propose that the true physical crack area can be accurately approximated as twice the
gradient-dependent energy. This leads to a new correction factor derived from energy quantities
already computed during the simulation. The proposed method is computationally efficient, applicable
to both 2D and 3D geometries without modification, and dynamically adapts to the evolving
phase-field. In addition, this correction eliminates the peak force observed during crack initiation in phase-field simulations, aligning the numerical results more closely with experimental observations and producing a convergent response.

This article is organized as follows. Section~\ref{sec:theory_and_models} summarizes the phase-field
fracture formulation and its finite element implementation. It also analyzes the crack surface
density functional, presents the energy equipartition principle, and provides an analytical
one-dimensional solution to illustrate this aspect. Section~\ref{sec:strain_localization}
analyzes the strain localization effect and its influence on crack area overestimation.
Section~\ref{sec:correction_methods} introduces the proposed
Double Gradient Correction Method (DGCM), comparing its performance with existing
element-based and skeletonization techniques. Section~\ref{sec:examples} validates the
proposed framework through several benchmarks, including complex curvilinear crack paths.
Finally, Section~\ref{sec:conclusions} summarizes the main findings, and
Section~\ref{sec:code_availability} provides details on the open-source software employed.

\section{Theory and models}
\label{sec:theory_and_models}
This section establishes the theoretical background for the proposed crack area correction method.
It reviews the main components of phase-field fracture theory, its discretization, the energy-controlled method for tracing equilibrium paths, the diffuse representation of cracks, and the equipartition of crack surface energy components, together with a corresponding one-dimensional illustration.

\subsection{Phase-field fracture model}
\label{sec:phase_field_fracture_model}
The phase-field fracture approach \cite{phase_field_FrancfortMarigo1998, phase_field_Miehe2010}
models cracks using a continuous scalar field, here denoted as $\phi$, which smoothly transitions from $\phi = 0$ in undamaged regions to $\phi = 1$ in fully damaged zones. This regularization replaces the sharp discontinuities of classical fracture mechanics with a diffuse interface whose width is governed by a length-scale parameter~$l$. As a result, the method can naturally capture complex crack initiation, branching, and merging without requiring the explicit tracking of crack surfaces or remeshing. The phase-field framework thus provides a unified and robust approach that can be used, after discretization, for simulating arbitrary crack geometries in both two and three dimensions.

In what follows, the behavior of the system is described by a total potential energy functional, $\mathcal{E}$, which consists of the bulk elastic energy and the energy dissipated by fracture:
\begin{equation}
  \mathcal{E}(\boldsymbol{u}, \phi) =
  \int_\Omega g(\phi) \psi(\boldsymbol{\epsilon}(\boldsymbol{u})) \,\mathrm{d}\Omega + G_c \int_\Omega \gamma(\phi, \nabla\phi) \,\mathrm{d}\Omega - \int_{\partial_N\Omega} \boldsymbol{t} \cdot \boldsymbol{u} \,\mathrm{d}S.
\label{eq:phase_field_fracture_functional}
\end{equation}
In this expression, $\boldsymbol{u}$ is the displacement field, and
$\boldsymbol{\epsilon}(\boldsymbol{u}) = \frac{1}{2}(\nabla \boldsymbol{u} + \nabla
\boldsymbol{u}^T)$ is the strain tensor. The first integral accounts for the elastic strain energy,
where $\psi(\boldsymbol{\epsilon}) = \frac{1}{2}\lambda (\text{tr}(\boldsymbol{\epsilon}))^2 + \mu
\boldsymbol{\epsilon}:\boldsymbol{\epsilon}$ is the strain energy density, with $\lambda$ and $\mu$
being the Lamé parameters. The material's stiffness is reduced by the degradation function $g(\phi)
= (1 - \phi)^2$. The second integral corresponds to the fracture energy, with $G_c$ being the
critical energy release rate and $\gamma(\phi, \nabla\phi) = \frac{1}{2l}\phi^2 + \frac{l}{2}
|\nabla \phi|^2$ representing the crack surface density function for the AT2 model \cite{phase_field_Miehe_lh_relation}, regularized by the length scale~$l$. The final term accounts for the work done by external tractions $\boldsymbol{t}$ on the Neumann boundary $\partial_N\Omega$; body forces are not considered in this study.

The governing equations are derived from the stationarity of the energy functional, $\delta\mathcal{E}=0$, with respect to the displacement and phase-field variables. This yields the following weak form:
\begin{equation}
   \int_\Omega g(\phi)\boldsymbol{\sigma}(\boldsymbol{\epsilon}(\boldsymbol{u})):\boldsymbol{\epsilon}(\delta \boldsymbol{u}) \,\mathrm{d}\Omega - \int_{\partial_N\Omega} \boldsymbol{t} \cdot \delta\boldsymbol{u} \,\mathrm{d}S  = {0} \ ,
\end{equation}
\begin{equation}
   \int_\Omega g'(\phi) \delta\phi \, \psi(\boldsymbol{\epsilon}(\boldsymbol{u})) \,\mathrm{d}\Omega + G_c \int_\Omega \left( \frac{1}{l} \phi \delta\phi  + l \nabla\phi \cdot \nabla \delta \phi \right) \,\mathrm{d}\Omega = 0\ ,
\end{equation}
where $\boldsymbol{\sigma}(\boldsymbol{\epsilon}) = \lambda (\text{tr}(\boldsymbol{\epsilon})) \boldsymbol I + 2\mu \boldsymbol{\epsilon}$ is the stress tensor, $g'(\phi)=-2(1-\phi)$ is the derivative of the degradation function with respect to $\phi$, and $\delta\boldsymbol{u}$ and $\delta\phi$ are admissible variations.

\subsubsection{Finite element discretization and solution strategy}
\label{sec:numerical_aspects}
The governing equations of the fracture model are discretized using the finite element method. Both the displacement field $\boldsymbol u$ and the phase-field $\phi$ are approximated with a standard Galerkin scheme using the same set of nodal shape functions $N_a$. The function space for the unknowns is defined as:
\begin{gather}
   V = \left\{ \nu: \Omega \rightarrow \mathbb{R},\ \nu(\boldsymbol x) = \sum_{a=1}^{n_{\text{node}}} N_a(\boldsymbol x)\, \nu_a \right\}
\end{gather}
The continuous fields and their variations are then expressed as:
\begin{gather}
   \boldsymbol u_h(\boldsymbol x) = \sum_{a=1}^{n_{\text{node}}} N_a(\boldsymbol x) \boldsymbol u_a, \quad \delta \boldsymbol u_h(\boldsymbol x) = \sum_{a=1}^{n_{\text{node}}} N_a(\boldsymbol x) \delta \boldsymbol u_a\ , \label{eq:fem_u} \\
   \phi_h(\boldsymbol x) = \sum_{a=1}^{n_{\text{node}}} N_a(\boldsymbol x) \phi_a, \quad \delta \phi_h(\boldsymbol x) = \sum_{a=1}^{n_{\text{node}}} N_a(\boldsymbol x) \delta \phi_a\ . \label{eq:fem_phi}
\end{gather}

Substituting these approximations into the weak form equations yields a system of non-linear algebraic equations. The arbitrariness of the variations $\delta \boldsymbol u_a$ and $\delta \phi_a$ requires the residual vectors at each node, $\boldsymbol{R}^{\boldsymbol u}_a$ and $R^{\phi}_a$, to be zero. These are defined as:
\begin{equation}
\label{eq-rurphi}
        \boldsymbol{R}^{\phi}_a = 
        \underset{e}{\mathbf{A}} \left( R^{\phi e}_a \right) = \boldsymbol{0} \quad \text{and} \quad 
        \boldsymbol{R}^{\boldsymbol u}_a = \underset{e}{\mathbf{A}} \left( R^{\boldsymbol u e}_a \right) = \boldsymbol{0}
\end{equation}
where $\mathbf{A}$ is the assembly operator. We refer to standard references (e.g. \cite{hughes1987vn,ern2004wx}) for the details of these manipulations. 

To robustly trace the complete equilibrium path during crack propagation, which is essential for analyzing the overestimation of the crack area, this work employs the non-variational energy-controlled scheme presented in \cite{phase_field_Castillon2025}. The weak form for this model is given by:
\begin{align}
   \int_\Omega g(\phi)\boldsymbol{\sigma}(\boldsymbol{\epsilon}(\boldsymbol{u})):\boldsymbol{\epsilon}(\delta \boldsymbol{u}) \,\mathrm{d}\Omega 
   - (1 - \Lambda c_2) \int_{\partial_N\Omega} \boldsymbol{t} \cdot \delta\boldsymbol{u} \,\mathrm{d}S &= {0}, \label{eq:non_variational_momentum} \\
   \int_\Omega g'(\phi) \delta\phi \, \psi(\boldsymbol{\epsilon}(\boldsymbol{u})) \,\mathrm{d}\Omega 
   + G_c \int_\Omega \left( \frac{1}{l} \phi \delta\phi + l \nabla\phi \cdot \nabla \delta \phi \right) \,\mathrm{d}\Omega &= 0, \label{eq:non_variational_phase_field} \\
   \delta \Lambda \left[ c_1 \int_\Omega \left( \frac{1}{2l} \phi^2 + \frac{l}{2} |\nabla \phi|^2 \right) \,\mathrm{d}\Omega 
   +  c_2 \int_{\partial_N\Omega} \boldsymbol{t} \cdot \boldsymbol{u} \,\mathrm{d}S - \tau(t) \right] &= 0.
   \label{eq:non_variational_constraint}
\end{align}
In these equations, $\Lambda$ is a scalar unknown, obtained by solving the problem such that the constraint in Eq.~\eqref{eq:non_variational_constraint} is satisfied. The vector $\boldsymbol{t}$ defines the direction of the applied traction, while its magnitude is governed by $\Lambda$. The constants $c_1$ and $c_2$ are introduced to ensure dimensional homogeneity in the constraint equation and to enhance numerical stability. The parameter $c_1$ has units of energy per unit length, consistent with the critical energy release rate $G_c$, while $c_2$ is dimensionless. This ensures that the control parameter $\tau(t)$ represents an energy quantity.

The resulting nonlinear system of equations is solved using an iterative method such as Newton-Raphson, where the scalar $\Lambda$ is treated as an additional global unknown.

Regarding the irreversibility of crack growth, the scalar $\Lambda$ enforces a monotonically
increasing function $\tau(t)$, ensuring that the weighted sum of the two energy terms in the constraint always increases over time. This condition still allows cases where, in multi-crack specimens, one crack decreases while another grows, or where energy redistributes between different regions. Nevertheless, this behavior has no impact on the solution strategy or the analyses carried out.

\subsection{Crack surface density functional}
\label{sec:crack_surface_density_functional}
The crack surface density functional is a key component of the phase-field model, as it approximates the crack's surface area. A detailed analysis of this functional is essential for two reasons. First, it governs the diffuse representation of the crack topology. Second, it provides the theoretical foundation for the correction method proposed in this work. 

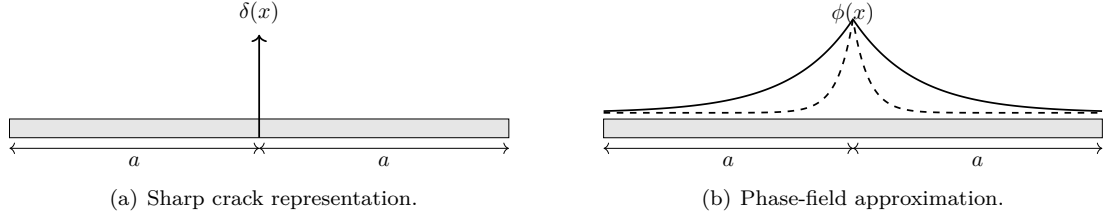
\begin{figure}[h!]
   \centering
   \subfigure[Sharp crack representation.]{
      \resizebox{0.45\textwidth}{!}{%
         \begin{tikzpicture}[scale=1.0]
            \draw[fill=gray!20, draw=black] (-4, -0.15) rectangle (4, 0.15); 

            \draw[thick, color=black] (0, -0.15) -- (0, 0.15); 

            \draw[<->] (-4, -0.32) -- (0, -0.32);
            \node[below] at (-2, -0.32) {$a$};
            \draw[<->] (0, -0.32) -- (4, -0.32);
            \node[below] at (2, -0.32) {$a$};

            \draw[thick, black, ->] (0, 0.15) -- (0, 1.5);
            \node[above, black] at (0, 1.55) {$\delta(x)$};
         \end{tikzpicture}%
      }
      \label{fig:CSDF_bar_sharp_crack}
   }
   \hfill
   \subfigure[Phase-field approximation.]{
      \resizebox{0.45\textwidth}{!}{%
         \begin{tikzpicture}[scale=1.0]
            \draw[fill=gray!20, draw=black] (-4, -0.15) rectangle (4, 0.15); 

            \draw[<->] (-4, -0.32) -- (0, -0.32);
            \node[below] at (-2, -0.32) {$a$};
            \draw[<->] (0, -0.32) -- (4, -0.32);
            \node[below] at (2, -0.32) {$a$};

            \draw[thick, black, domain=-4:4, samples=1000, smooth] 
            plot (\x, {1.5*exp(-abs(\x)/1.0) + 0.25});

            \draw[thick, black, dashed, domain=-4:4, samples=1000, smooth] 
                        plot (\x, {1.5*exp(-abs(\x)/0.25) + 0.25});

            \node[above, black] at (0, 1.55) {$\phi(x)$};
         \end{tikzpicture}%
      }
      \label{fig:CSDF_bar_phase_field_crack}
   }
   \caption{Bar with a central crack: (a) Sharp crack representation as a discontinuity at the center. (b) Phase-field approximation showing the smooth transition of $\phi(x)$, regularized by the length scale~$l$.}
   \label{fig:CSDF_bar_with_central_crack}
\end{figure}

Figure~\ref{fig:CSDF_bar_with_central_crack} illustrates the concept of crack representation using a simple one-dimensional bar with a central crack. In Figure~\ref{fig:CSDF_bar_sharp_crack}, the sharp crack is represented as a discontinuity at the center of the bar. Figure~\ref{fig:CSDF_bar_phase_field_crack} shows the phase-field approximation for two different length scale values, where the crack is modeled by a smooth transition of the phase-field variable $\phi(x)$, regularized by the length scale parameter~$l$. As the length scale value decreases, the crack approximation converges to the sharp representation. This demonstrates how the phase-field method provides a continuous approximation of the crack in a one-dimensional setting.

The term multiplying the critical energy release rate $G_c$ in Eq.~\eqref{eq:phase_field_fracture_functional} is the crack surface functional, which approximates the total crack area. This functional, denoted as $\Gamma[\phi]$, is defined for the AT2 model as:
\begin{equation}
   \Gamma[\phi] = \int_\Omega \gamma(\phi, \nabla \phi) \, d\Omega = \int_\Omega \left( \frac{1}{2l} \phi^2 + \frac{l}{2} |\nabla \phi|^2 \right) d\Omega.
   \label{eq:general_phase_field}
\end{equation}
The product $G_c \Gamma[\phi]$ represents the total fracture energy, which has units of energy as expected. For the sake of simplicity in the subsequent analysis, we will normalize the fracture energy by setting $G_c=1$ and refer to $\Gamma[\phi]$ as the crack surface energy.

This functional is designed to be $\Gamma$-convergent, ensuring that as the length scale parameter $l \to 0$, the diffuse crack representation converges to a sharp crack, and $\Gamma[\phi]$ correctly approximates the true surface energy.

For the purpose of analysis, it is useful to decompose $\Gamma[\phi]$ into two distinct energy
contributions: a term dependent explicitly on the phase-field variable, $\Gamma_{\phi}$, and a term dependent on its gradient, $\Gamma_{\nabla \phi}$:
\begin{align}
   \Gamma_{\phi}[\phi] &= \int_\Omega \frac{1}{2l} \phi^2 \, d\Omega, \label{eq:gamma_phi_def} \\
   \Gamma_{\nabla \phi}[\phi] &= \int_\Omega \frac{l}{2} |\nabla \phi|^2 \, d\Omega. \label{eq:gamma_gradphi_def}
\end{align}
The equilibrium equation is derived from the stationarity condition $\delta \Gamma = 0$, leading to the variational form:
\begin{equation}
    \Gamma'[\phi] := \int_\Omega \left(\frac{1}{l}\phi\delta \phi + l \nabla\phi \cdot \nabla\delta\phi \right) \, d\Omega = 0.
    \label{eq:variational_form_gamma}
\end{equation}

\subsection{Equipartition of the fracture energy}
\label{sec:equipartition}
A key result for the remainder of the article is the \emph{equipartition} of the fracture energy. This result, which is proven next, says that the two terms of the fracture energy contribute equally in the limit $l\to0$. For simplicity, we focus on the three-dimensional problem with the AT2 model.

To start, let us note that the phase-field function $\phi$ belongs to $H^1(\Omega)$, the space of functions defined on the domain $\Omega$ that are square-integrable and have square-integrable (weak) derivatives. These functions satisfy Young's inequality
\begin{equation}
\label{eq-young1}
\int_{\Omega}
|\phi| \, |\nabla\phi|\; d\Omega
\le
\int_{\Omega}
\left[
\frac{|\phi|^2}{2\epsilon^2} + \frac{\epsilon^2}{2} |\nabla\phi|^2 
 \right]\, d\Omega\ ,
\end{equation}
for all $\epsilon\ne0$. Moreover, the equality is verified only if $|\phi|/\epsilon=\epsilon|\nabla\phi|$ almost everywhere. Next, we observe that the left hand side of this inequality can be written as
\begin{equation}
\int_{\Omega}
|\phi| \, |\nabla\phi|\; d\Omega
=
\frac{1}{2}\int_{\Omega} |\nabla (\phi^2) |\; d\Omega\ ,
\end{equation}
and the right-hand integral is known as the \emph{total variation} of $\phi^2$.

Consider now a fracture problem with a crack $\mathcal{C}$ of area $|\mathcal{C}|$. The trace of the phase field $\phi$ would be 1 on the crack and 0 far from it. Then, define coordinates  $(\xi_1,\xi_2,\xi_3)$ for $\Omega$ such that the first two are surface coordinates of the crack and the third one is normal to $\mathcal{C}$. The set $\Omega$ can then be split into three regions: one with positive $\xi_3$ denoted $\Omega^+$; a second one, $\Omega^-$ with negative $\xi_3$, and a third one where the coordinates $(\xi_1,\xi_2,\xi_3)$ are not well-defined due to the singularity of the crack edge. Then, since the normal to the crack corresponds with the $\xi_3$ direction, we can evaluate a bound for the total variation of $\phi^2$ to be 
\begin{equation}
\label{eq-measure1}
\int_{\Omega} |\nabla(\phi^2)|\;d\Omega 
\ge
\int_{\Omega^+\cup \Omega^-} |\nabla(\phi^2)|\;d\Omega 
=
\int_{\mathcal{C}} \int 
\left|\frac{\partial}{\partial \xi_3} (\phi^2)\right| \;d\xi_3\,d\xi_1\,d\xi_2
= 2\, |\mathcal{C}|\ ,
\end{equation}
where the area of the crack surface must be counted twice since the integral must be performed on 
$\Omega^+$ and $\Omega^-$. From the Modica-Mortola theory, we know that the fracture energy satisfies
\begin{equation}
\label{eq-mm1}
\lim_{l\to 0} G_c \int_{\Omega}
\left[
\frac{\phi^2}{2l} + \frac{l}{2} |\nabla\phi|^2
 \right]\; d\Omega 
 =
 G_c\; |\mathcal{C}|\ .
\end{equation}
Then, using Eq.~\eqref{eq-young1} with $\epsilon=\sqrt{l}$, and Eqs.~\eqref{eq-measure1}-\eqref{eq-mm1}, it follows that
\begin{equation*}
G_c \; |\mathcal{C}|
=
\lim_{l\to 0}
G_c
\int_{\Omega}
\left[ 
\frac{\phi^2}{2l} + \frac{l}{2} |\nabla\phi|^2 
 \right]\; d\Omega\
\ge
\frac{G_{c}}{2}\int_{\Omega} |\nabla(\phi^2)| \, d\Omega
\ge
G_c\,|\mathcal{C}|\ .
\end{equation*}
For these identities to hold, the inequalities must be equalities, which occurs only if
\begin{equation}
\lim_{l\to 0}
\int_{\Omega} \frac{\phi^2}{2l} \;d\Omega = 
\lim_{l\to 0}
\int_{\Omega} \frac{l}{2}|\nabla\phi|^2 \;d\Omega,
\end{equation}
which is precisely the equipartition result.

We note that the equipartition of the fracture energy only happens in the limit $l\to0$ when the Modica-Mortola functional converges to the crack area and the total variation of $\phi^2$ can be evaluated.

\subsubsection{One-dimensional case}
\label{sec:one_dimensional_analytical_solution}
Here, a one-dimensional problem for which the analytical solution can be obtained is presented to illustrate the principle of energy equipartition. We consider a bar with a central crack, as illustrated in Figure~\ref{fig:CSDF_bar_with_central_crack}. The equilibrium profile of the phase-field variable, $\phi(x)$, is governed by the following ordinary differential equation (in the weak sense):
\begin{equation}
   \frac{1}{l}\phi(x) - l \phi''(x) = 0,
   \label{eq:ode_phase_at2_1d}
\end{equation}
subject to the boundary conditions $\phi(0) = 1$ and, due to symmetry, $\phi'(\pm a) = 0$. The analytical solution is given by:
\begin{equation}
   \phi(x) = e^{-|x|/l} + \frac{2 \sinh \left( \frac{|x|}{l} \right)}{e^{\frac{2a}{l}}+1},
\end{equation}
with its corresponding derivative:
\begin{equation}
   \phi'(x) = \frac{-\text{sign}(x)}{l} e^{-|x|/l} + \frac{2\,\text{sign}(x)}{l} \frac{\cosh\left(\frac{|x|}{l}\right)}{e^{\frac{2a}{l}} + 1},
\end{equation}
where $\text{sign}(x)$ is the sign function, defined as $1$ for $x > 0$, $-1$ for $x < 0$, and $0$
for $x = 0$. As the ratio $a/l \to \infty$, this solution converges to the well-known infinite
domain solution, $\phi(x) = e^{-|x|/l}$ (see \cite{phase_field_Miehe2010}).

\begin{figure}[h!]
   \centering
   \subfigure[Phase-field profile $\phi(x)$ for various length scales.]{
      \includegraphics[width=0.45\textwidth]{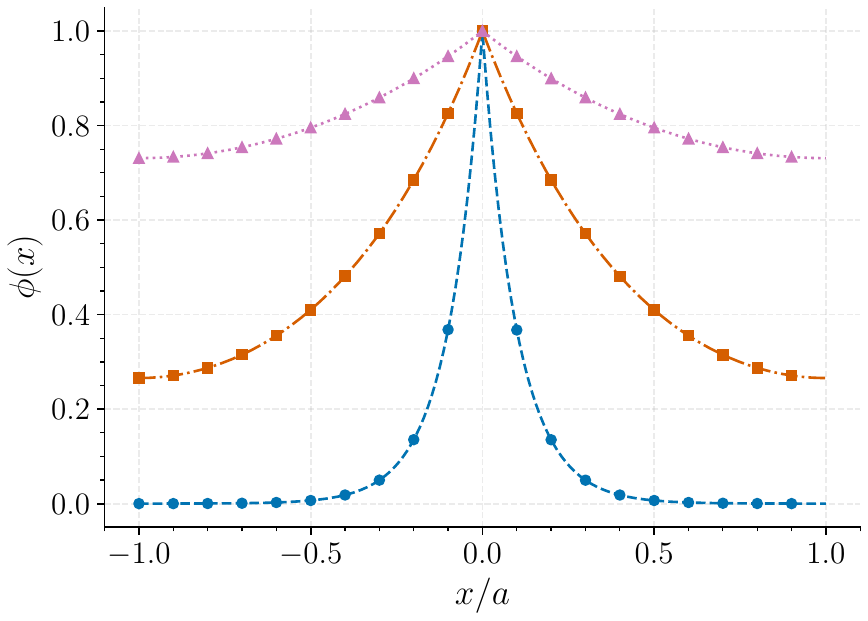}
      \label{fig:CSDF_theory_phase_field_profile_at2}
   }
   \hfill
   \subfigure[Phase-field derivative profile $\phi'(x)$.]{
      \includegraphics[width=0.45\textwidth]{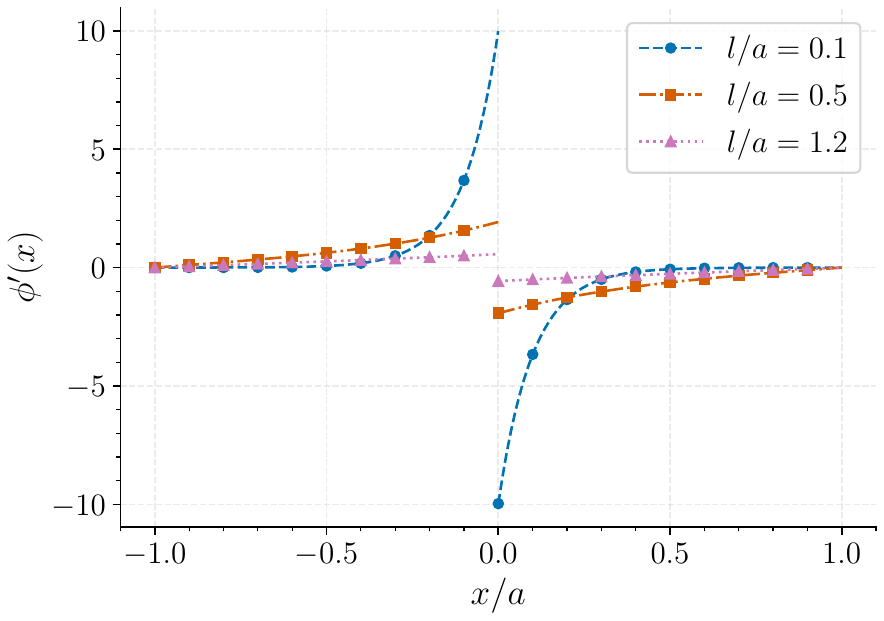}
      \label{fig:CSDF_theory_phase_field_grad_profile_at2}
   }
   \caption{Comparison of analytical solutions for the diffuse crack representation. (a) Phase-field profile $\phi(x)$ and (b) its derivative $\phi'(x)$ for various length scales $l$.}
   \label{fig:CSDF_theory_phase_field_length_scale}
\end{figure}

\begin{figure}[h!]
   \centering
   \includegraphics[width=14cm]{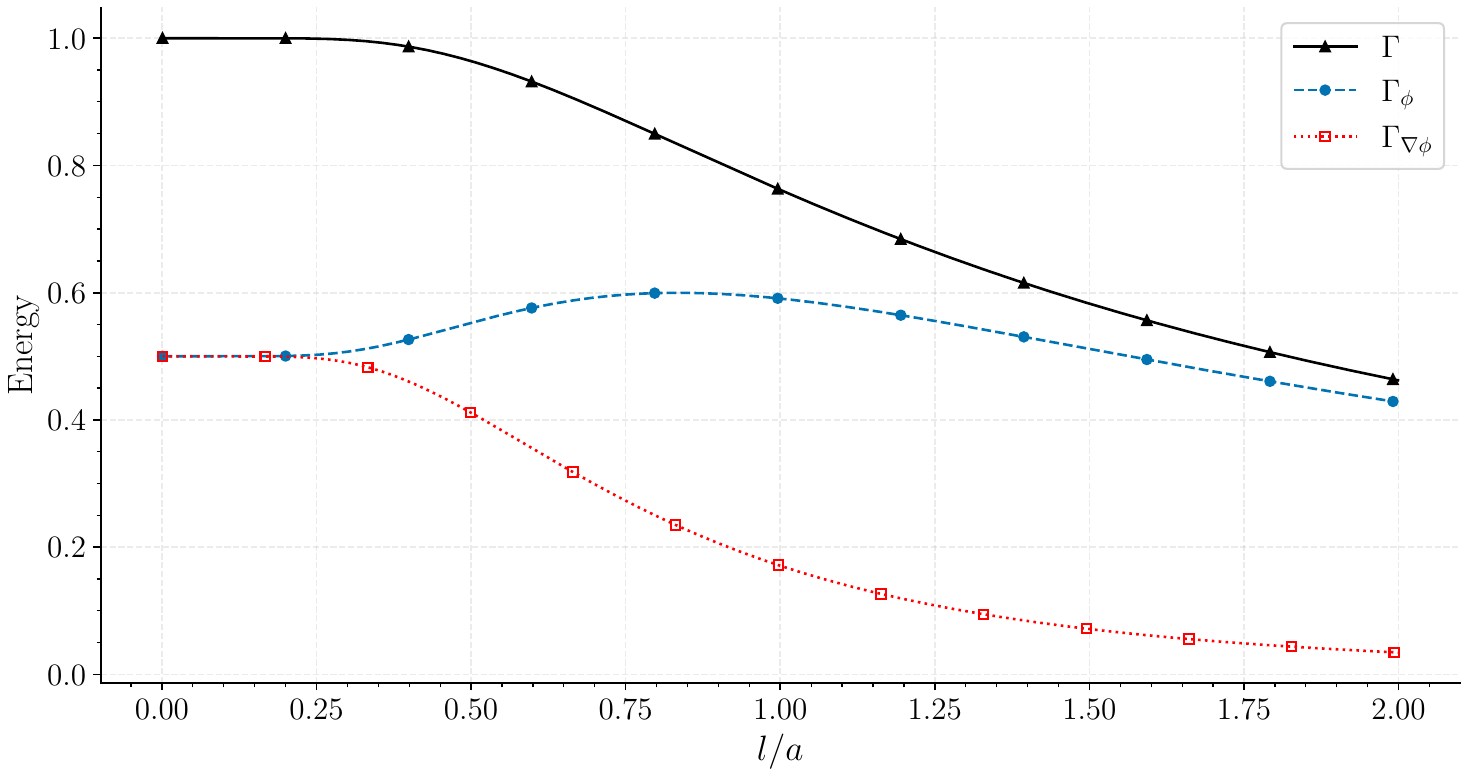}
   \caption{Energy as a function of the $l/a$ ratio. As $l/a$ decreases, the total energy approaches one, and the contributions from the phase-field and its gradient become equal.}
   \label{fig:CSDF_phase_field_at2_energy_l_vs_energy}
\end{figure}

Figure~\ref{fig:CSDF_theory_phase_field_length_scale} shows the analytical phase-field profile $\phi(x)$ and its derivative $\phi'(x)$ for different length scales. As $l$ decreases, the profile sharpens and the boundary effects diminish, approaching the behavior of a discrete crack.
The total energy $\Gamma$ can be decomposed into two contributions, as defined in Eqs.~\eqref{eq:gamma_phi_def} and \eqref{eq:gamma_gradphi_def}: the phase-field term, $\Gamma_{\phi}$, and the gradient term, $\Gamma_{\nabla \phi}$. Substituting the analytical solution into these definitions yields:
\begin{align}
   \Gamma_{\phi} &= \int_{-a}^{a} \frac{1}{2l}\phi(x)^2\, dx = \frac{1}{2} \tanh\left(\frac{a}{l}\right) + \frac{a}{2l} \left[1-\tanh^2\left(\frac{a}{l}\right)\right], \label{eq:gamma_phi} \\
   \Gamma_{\nabla \phi} &= \int_{-a}^{a} \frac{l}{2}(\phi'(x))^2\, dx = \frac{1}{2} \tanh\left(\frac{a}{l}\right) - \frac{a}{2l} \left[1-\tanh^2\left(\frac{a}{l}\right)\right]. \label{eq:gamma_gradphi}
\end{align}
The total energy is the sum of these terms, which simplifies to:
\begin{equation}
   \Gamma = \Gamma_{\phi} + \Gamma_{\nabla \phi} = \tanh\left(\frac{a}{l}\right).
   \label{eq:total_energy_at2}
\end{equation}
Figure~\ref{fig:CSDF_phase_field_at2_energy_l_vs_energy} illustrates the total energy and its individual components as a function of the ratio $l/a$. As this ratio approaches zero ($l/a \to 0$), the total energy converges to one, and the contributions from the phase-field and gradient terms become equal, such that $\Gamma_{\phi} = \Gamma_{\nabla \phi}$. This illustrates energy equipartition in the one-dimensional case, in accordance with the theoretical result derived in the previous section.

\subsubsection{Finite element error in the one dimensional problem}
\label{sec:fem_error_one_dimensional_problem}
When minimizing the crack surface density functional, it is crucial to ensure that the element size is sufficiently small to accurately resolve the phase-field profile. If the element size is too large relative to the length-scale parameter~$l$, the numerical solution may fail to adequately capture the phase-field and its gradient. This leads to inaccuracies in the representation of the crack surface and its associated energy. A common guideline \cite{phase_field_Miehe_lh_relation}, particularly for linear elements, is to maintain a ratio of $l/h \ge 2$.

To illustrate this, the one-dimensional problem presented in Section~\ref{sec:one_dimensional_analytical_solution} is solved using the finite element method with linear elements, following the discretization in Eq.~\eqref{eq:fem_phi} and the approach in \cite{phase_field_effective_Gc_factor_2}. Exploiting symmetry, only half of the bar is modeled, with a Dirichlet boundary condition of $\phi=1$ applied at the symmetry plane to represent the central crack. We consider parameters $l=0.1$~mm and $a=1.0$~mm, resulting in a ratio of $a/l=10$. In this configuration, boundary effects are negligible, and the analytical energy values derived from Eqs.~\eqref{eq:gamma_phi}, \eqref{eq:gamma_gradphi}, and \eqref{eq:total_energy_at2} can be treated as the infinite domain solution: $\Gamma=1.0$, with $\Gamma_{\phi}=\Gamma_{\nabla \phi}=0.5$.

\begin{figure}[h!]
   \centering
   \includegraphics[width=8cm]{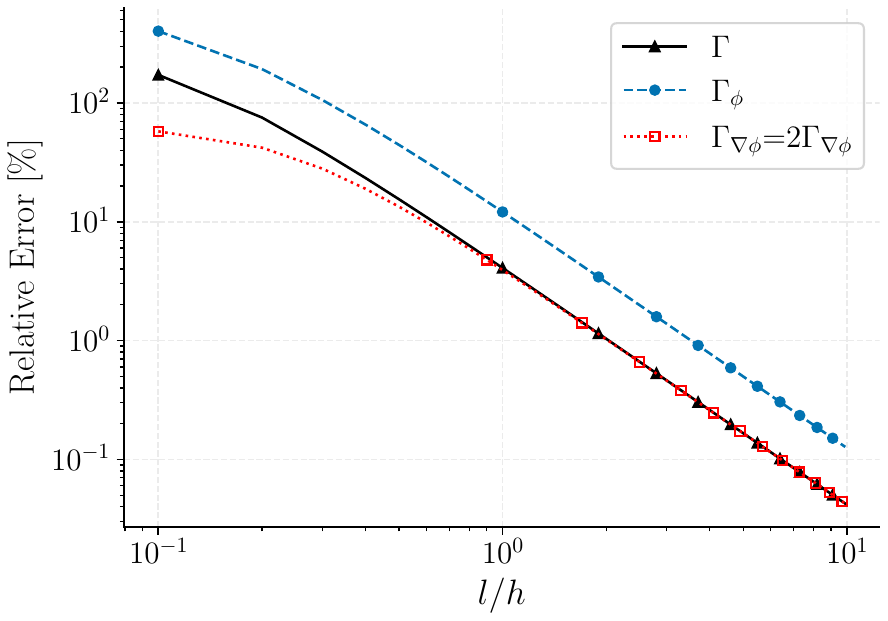}
   \caption{Relative error in the total crack surface energy $\Gamma$, phase-field energy component $\Gamma_{\phi}$, and gradient energy component $\Gamma_{\nabla \phi}$ in a finite element one dimensional simulation as a function of the mesh resolution ratio $l/h$.}
   \label{fig:crack_surface_fem_error}
\end{figure}

The relative error in the energy terms is evaluated for different element sizes, corresponding to $l/h$ ratios from 0.1 to 10. As shown in Figure~\ref{fig:crack_surface_fem_error}, the relative error for the phase-field dependent energy, $\Gamma_{\phi}$, is significantly higher than the error for the gradient-dependent energy, $\Gamma_{\nabla \phi}$. It is also observed that for a ratio of $l/h \approx 2.0$, the relative error in the total energy is approximately $1\%$.

Note that in this scenario, where boundary effects are negligible and the energy equipartition result holds ($\Gamma_{\phi}=\Gamma_{\nabla \phi}$), the total energy can be approximated as twice the gradient energy ($2\Gamma_{\nabla \phi}$). By applying this approximation, the error in the total energy becomes equivalent to the error in the gradient energy term, which is lower than the error obtained by directly summing the two numerical components.

\section{Strain localization effect on crack area}
\label{sec:strain_localization}
\begin{figure}[h!]
   \centering
   \resizebox{0.5\textwidth}{!}{%
      \begin{tikzpicture}[scale=1.0]
         \draw[fill=gray!20, draw=black] (-4, -0.15) rectangle (4, 0.15); 
         \def\h{0.25};
         \draw[<->] (-4, -0.32) -- (0, -0.32);
         \node[below] at (-2, -0.32) {$a$};
         \draw[<->] (0, -0.32) -- (4, -0.32);
         \node[below] at (2, -0.32) {$a$};
         \draw[thick, dashed, black, domain=-4:4, samples=1000, smooth] 
            plot (\x, {1.5*exp(-abs(\x)/1.0) + \h});
         \draw[thick, black, domain=(-4.0+\h):-0.0, samples=1000, smooth] 
            plot (\x-\h, {1.5*exp(-abs(\x)/1.0) + \h});
         \draw[thick, black, domain=(-\h):(\h), samples=1000, smooth] 
            plot (\x, {1.75});
         \draw[thick, black, domain= 0.0:(4.0-\h), samples=1000, smooth] 
            plot (\x+\h, {1.5*exp(-abs(\x)/1.0) + \h});
         \node[above] at (0, 2.0) {$h$};
         \draw[<->] (-\h, 2.0) -- (\h, 2.0);
      \end{tikzpicture}%
   }
   \caption{Illustration of the phase-field profile distortion caused by strain localization. The ideal profile (dashed line) is artificially flattened to $\phi=1$ over an element of size $h$ (solid line), leading to an overestimation of the crack surface energy.}
   \label{fig:bar_with_central_crack_correction}
\end{figure}
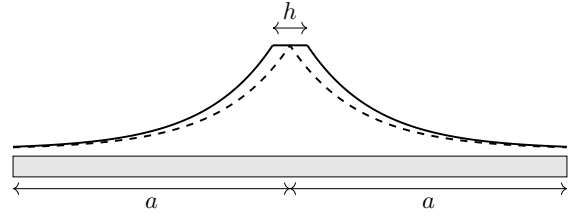

A common numerical artifact in phase-field fracture simulations is strain localization, as analyzed
in \cite{phase_field_effective_Gc_factor_2}. In this phenomenon, the phase-field variable
artificially saturates to $\phi=1$ across an entire element of size $h$, leading to an
overestimation of the crack surface area, when the latter is measured from the fracture energy functional. Figure~\ref{fig:bar_with_central_crack_correction} illustrates how the ideal phase-field profile is distorted by this effect.

Within the localized element, the phase-field is constant ($\phi=1$), and its gradient is zero. Figure~\ref{fig:phase_field_strain_location_h} and \ref{fig:phase_field_gradient_strain_location_h} illustrate this for the phase-field and its gradient, respectively. This alters the energy calculation. The total energy, taking the strain localization effect, is denoted as $\Gamma_{\text{sl}}$, and can be obtained by modifying the integral in Eq.~\eqref{eq:general_phase_field} to account for the localized region:

\begin{figure}[h!]
   \centering
   \subfigure[Phase-field profile $\phi(x)$ with strain localization.]{
     \includegraphics[width=0.45\textwidth]{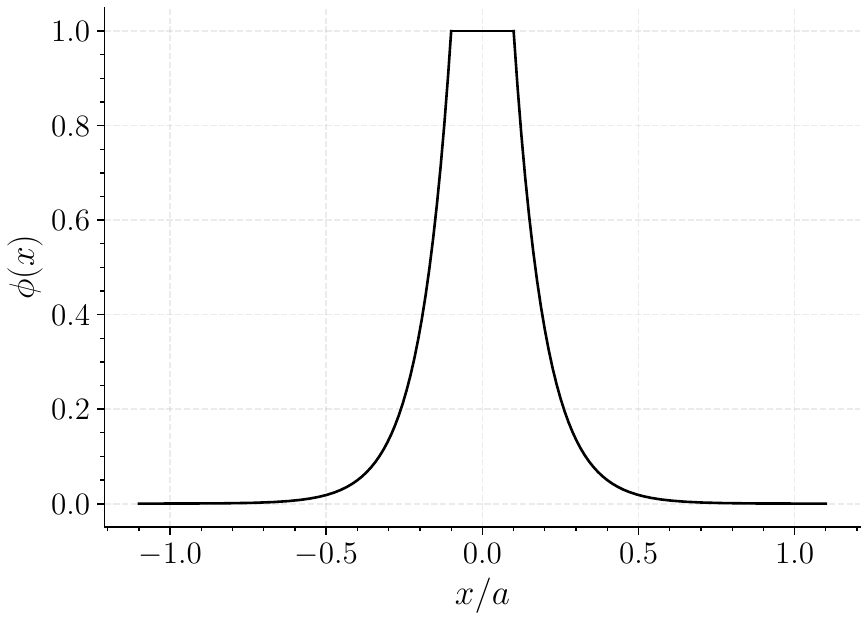}
     \label{fig:phase_field_strain_location_h}
   }
   \hfill
   \subfigure[Phase-field gradient profile $\phi'(x)$ with strain localization.]{
     \includegraphics[width=0.45\textwidth]{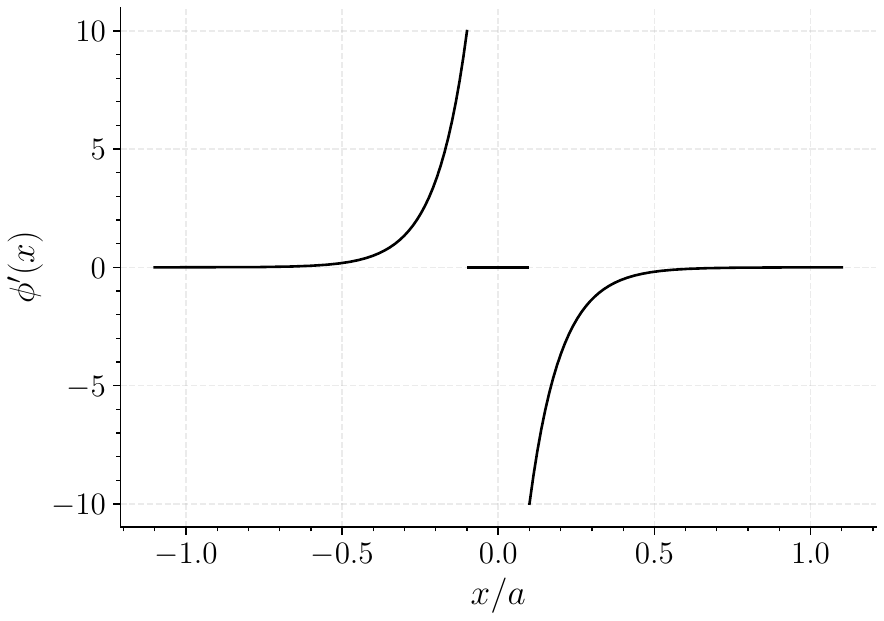}
     \label{fig:phase_field_gradient_strain_location_h}
   }
   \caption{Effect of strain localization on the phase-field profile and its gradient for a ratio $l/a=0.1$. The profiles exhibits a plateau of width $h$ where $\phi=1$ and $\phi'=0$, deviating from the ideal analytical profile.}
   \label{fig:strain_location_h}
\end{figure}

\begin{equation}
   \Gamma_{\text{sl}}[\phi, \nabla \phi] = \int_{-a}^{+a}  \left( \frac{1}{2l} \phi^2 + \frac{l}{2} (\phi')^2 \right) dx + \int_{0}^{h}  \left( \frac{1}{2l} (1)^2 + \frac{l}{2} (0)^2 \right) dx.
   \label{eq:strain_localization_energy_1d}
\end{equation}
As a result, the energy contributions are modified. The phase-field term, $\Gamma_{\phi}$, increases due to the contribution from the localized region, whereas the gradient term, $\Gamma_{\nabla\phi}$, remains unchanged because the gradient vanishes where $\phi$ is constant:
\begin{align}
   \Gamma_{\phi, \text{sl}} &=  \frac{1}{2} \tanh\left(\frac{a}{l}\right) + \frac{a}{2l} \left[1-\tanh^2\left(\frac{a}{l}\right)\right] + \frac{h}{2l}, \\
   \Gamma_{\nabla\phi, \text{sl}} & = \Gamma_{\nabla\phi} = \frac{1}{2} \tanh\left(\frac{a}{l}\right) - \frac{a}{2l} \left[1-\tanh^2\left(\frac{a}{l}\right)\right].
\end{align}
The total energy with strain localization becomes:
\begin{equation}
   \Gamma_{\text{sl}} = \Gamma_{\phi, \text{sl}} + \Gamma_{\nabla\phi, \text{sl}} = \tanh\left(\frac{a}{l}\right) + \frac{h}{2l}.
\end{equation}
This shows that strain localization introduces an additive error of $h/(2l)$ to the total crack surface energy in the one dimensional case. This error is directly proportional to the element size $h$ and inversely proportional to the length scale~$l$.

\begin{figure}[h!]
   \centering
   \subfigure[Total energy $\Gamma_{\text{sl}}$ vs. $l/a$ ratio.]{
     \includegraphics[width=0.45\textwidth]{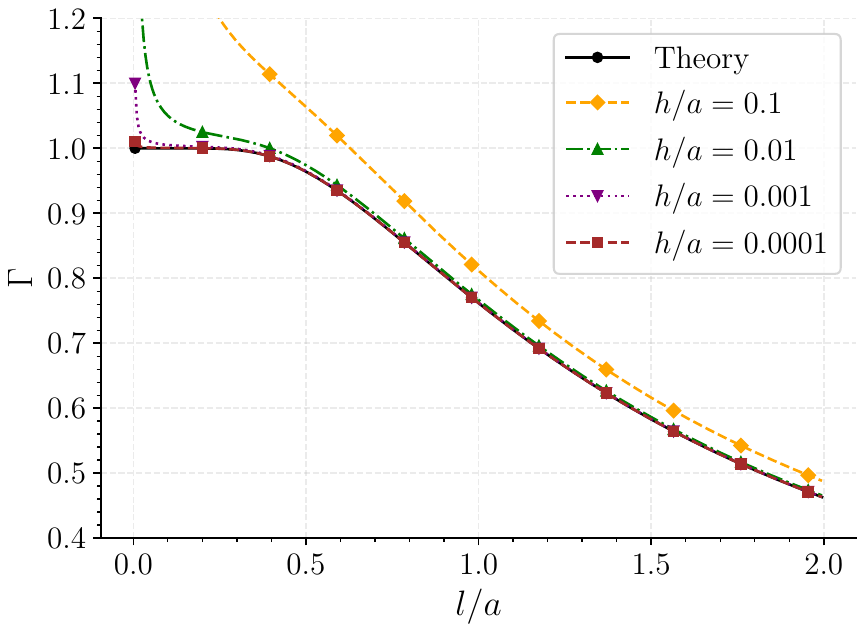}
     \label{fig:energy_strain_location_h}
   }
   \hfill
   \subfigure[Phase-field energy $\Gamma_{\phi, \text{sl}}$ vs. $l/a$ ratio.]{
     \includegraphics[width=0.45\textwidth]{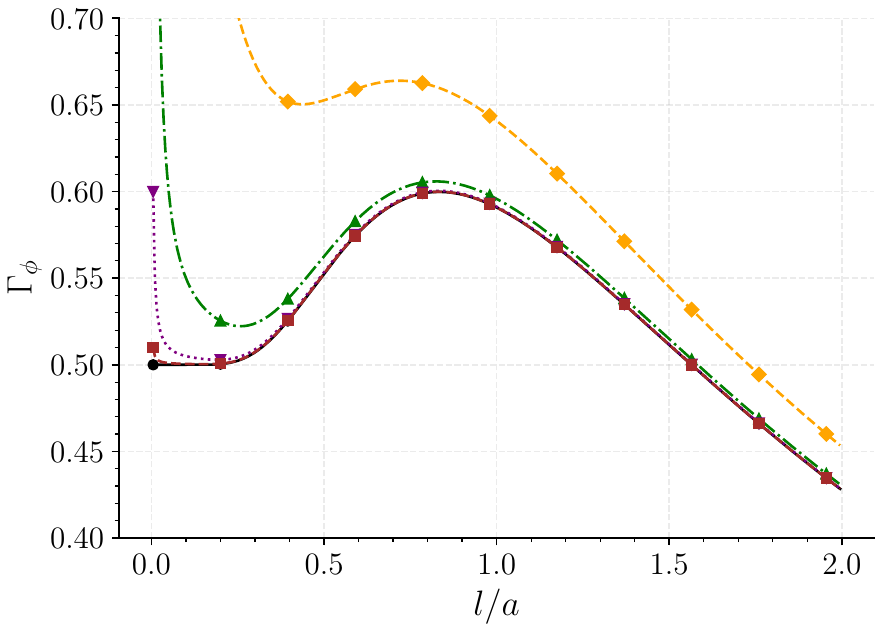}
     \label{fig:energy_phi_strain_location_h}
   }
   \caption{Effect of strain localization on energy contributions for different element sizes $h$. As $h$ decreases, both the total energy (a) and the phase-field energy (b) converge to the ideal theoretical solution (solid black line).}
   \label{fig:energystrain_location_h}
\end{figure}

Figures~\ref{fig:energy_strain_location_h} and \ref{fig:energy_phi_strain_location_h} illustrate the impact of strain localization on the computed energies. These plots show the total energy and the phase-field energy component, respectively, for various $h/a$ ratios. As the element size $h$ decreases, the additional energy introduced by strain localization diminishes, and the computed energies converge toward the theoretical values. In contrast, the gradient-dependent energy component remains unaffected by this artifact and therefore coincides with the theoretical $\Gamma_{\nabla\phi}$ curve shown in Figure~\ref{fig:CSDF_phase_field_at2_energy_l_vs_energy}.

\section{Crack surface overestimation correction}
\label{sec:correction_methods}
Accurately measuring the crack surface area is a critical challenge in phase-field modeling. The standard approach of integrating the crack surface density functional, $\Gamma[\phi]$, over the domain typically overestimates the true physical crack area. This discrepancy arises from numerical artifacts like strain localization, which artificially inflates the computed crack surface. As demonstrated in \cite{phase_field_Castillon2025}, this overestimation systematically affects all related simulation quantities—including the forces, displacements, and energies—compromising the physical accuracy of the results even when the numerical solution correctly satisfies the governing equations. To address this issue, several correction methods have been developed.

Following the approach presented in \cite{phase_field_Castillon2025}, the crack area $\Gamma$, defined as the area that would be obtained in the absence of strain localization, is determined by scaling the simulated area $\Gamma_\mathrm{sl}$ (which includes the energy overestimation due to strain localization) by a correction factor $\mathcal{F}$:
\begin{equation}
   \Gamma = \frac{\Gamma_\mathrm{sl}}{\mathcal{F}}.
   \label{eq:crack_area_correction}
\end{equation}
Here, $\mathcal{F}$ is a factor that quantifies the overestimation caused by the diffuse crack
representation and numerical artifacts. 

\subsection{Physically consistent application of the correction factor}
\label{sec:consistent_application_factor}
To clarify and align with other approaches that handle this phenomenon, it is important to note that we consider the crack surface area to be overestimated. This overestimation can also be interpreted as an effectively higher energy release rate. However, it is crucial to distinguish whether the crack surface is overestimated or the critical energy release rate is effectively higher than expected. Several considerations can be taken into account to correct all quantities of the phase-field fracture problem while satisfying the governing equations. In this context, three alternative, mathematically consistent schemes (\textbf{Scheme I}, \textbf{Scheme II}, and \textbf{Scheme III}) for transforming simulated results into physical quantities using the correction factor $\mathcal{F}_\mathrm{corr}$ are presented in Table~\ref{tab:ch_f_factor_corrected_alternatives}. Each scheme applies the correction factor differently to the critical energy release rate, crack area, force, and displacement, while ensuring that the overall physical consistency of the model is maintained. Furthermore, the degraded strain energy, denoted as $\Psi(\boldsymbol{u}, \phi) = \int_\Omega g(\phi) \, \psi(\boldsymbol \epsilon(\boldsymbol{u})) \, \mathrm{d}\Omega$, and the specimen stiffness given as $K=P/F$, is also considered.
\begin{table}[h!]
   \centering
   \resizebox{0.9\textwidth}{!}{
   \renewcommand{\arraystretch}{1.5}
   \setlength{\tabcolsep}{12pt}
   \begin{tabular}{llll}
         \toprule
         \textbf{Physical Quantity} & \textbf{Scheme I} & \textbf{Scheme II} & \textbf{Scheme III} \\ 
         \midrule
         Critical energy release rate $G_{c,\mathrm{phys}}$ & $G_{c,\mathrm{sl}}$ & $G_{c,\mathrm{sl}}\mathcal{F}$ & $G_{c,\mathrm{sl}}\mathcal{F}$ \\
         Crack area $\Gamma_\mathrm{phys}$                 & $\Gamma_\mathrm{sl} / \mathcal{F}$ & $\Gamma_\mathrm{sl}$ & $\Gamma_\mathrm{sl} / \mathcal{F}$ \\
         Degraded Strain Energy ($\Psi$) & $\Psi_\mathrm{sl} / \mathcal{F}$
          & $\Psi_\mathrm{sl} \mathcal{F} $ & $ \Psi_\mathrm{sl}$ \\
         Force $P_\mathrm{phys}$                        & $P_\mathrm{sl} / \sqrt{\mathcal{F}}$ & $P_\mathrm{sl}\sqrt{\mathcal{F}}$ & $P_\mathrm{sl}$ \\
         Displacement $u_\mathrm{phys}$                 & $u_\mathrm{sl} / \sqrt{\mathcal{F}}$ & $u_\mathrm{sl} \sqrt{\mathcal{F}}$ & $u_\mathrm{sl}$ \\
         Stiffness ($K$) & $K_\mathrm{phys} = K_\mathrm{sl}$ & $K_\mathrm{phys} = K_\mathrm{sl}$ & $K_\mathrm{phys} = K_\mathrm{sl}$ \\
         \bottomrule
   \end{tabular}
   }
   \caption{Three alternative mathematically consistent schemes (\textbf{Scheme I}, \textbf{Scheme II}, and \textbf{Scheme III}) for transforming simulated results into physical quantities using the correction factor $\mathcal{F}$.}
   \label{tab:ch_f_factor_corrected_alternatives}
\end{table}
To emphasize these alternatives, the relationship for the energy release rate in the problem under consideration, 
\begin{equation}
G_c = \frac{P^2}{2} \frac{\partial C}{\partial a},
\label{eq:energy_release_rate}
\end{equation}
where $P$ is the applied force, $C$ is the structural compliance of the specimen, and $a$ is the crack area, can be expressed for each column as follows:
\begin{itemize}
   \item \textbf{Scheme I}: $G_c = \frac{(P / \sqrt{\mathcal{F}})^2}{2B} \left( \frac{\mathrm{d}C}{\mathrm{d}a} \cdot \mathcal{F} \right)$
   \item \textbf{Scheme II}: $G_c\mathcal{F} = \frac{(P\sqrt{\mathcal{F}})^2}{2B} \frac{\mathrm{d}C}{\mathrm{d}a}$
   \item \textbf{Scheme III}: $G_c\mathcal{F} = \frac{P^2}{2B} \left( \frac{\mathrm{d}C}{\mathrm{d}a} \cdot \mathcal{F} \right)$
\end{itemize}
Simplifying these expressions shows that the fundamental relationship \eqref{eq:energy_release_rate} is satisfied in all cases, confirming the thermodynamic consistency of the transformations. Another way to verify this aspect is by considering the dimensionless phase-field fracture scheme, where it can be shown that the PFF problem is defined by the Poisson's ratio $\nu$ and the dimensionless length scale parameter, as presented in \cite{phase_field_non_dimensional}.

As mentioned previously, the correction framework adopted in this work corresponds to \textbf{Scheme I}, motivated by the following considerations:
\begin{itemize}
   \item The critical energy release rate $G_c$ is an intrinsic material property; therefore, it should not be modified by a correction factor addressing a numerical artifact.
   \item Since the inaccuracy is directly related to the overestimation of the crack area, it is physically more intuitive to apply the correction to the crack length rather than the energy release rate.
   \item Force and displacement are global quantities that depend on the overall system response. Consequently, it is preferable to apply the correction only to the output quantities resulting from the simulations, rather than to the material input parameters, to accurately reflect the true physical behavior of the problem.
   \item In this manner, the exact physical parameter (the given $G_c$) is preserved, and the correction is systematically applied to the simulation outputs, namely the crack area, forces, and displacements.
\end{itemize}
However, it must be noted that all three cases are mathematically consistent. The choice of the correction scheme can ultimately depend on practical considerations, such as ease of implementation or the specific interpretability required for the results.

In the general literature, the correction is typically applied to $G_c$, which corresponds to \textbf{Scheme II} or \textbf{Scheme III}. However, this approach can lead to confusion, as modifying $G_c$ can alter the force or the crack length, but not both simultaneously. Furthermore, the effect on displacement and the overall mathematical consistency of the system are rarely addressed. Concurrently correcting the force using \textbf{Scheme II} and the crack area using \textbf{Scheme III} mixes assumptions and produces physically inconsistent results. For these reasons, the unambiguous methodology presented in \textbf{Scheme I} is preferred in this work.

For all cases, as the displacement and force are corrected by the same factor, the stiffness of the system remains unchanged.

Based on this, the preferred Scheme I yields the relative error introduced by the strain localization effect, which can be expressed in terms of the correction factor as:
\begin{equation}
   \text{Relative Error} = \frac{|\Gamma - \Gamma_\mathrm{sl}|}{|\Gamma|} = \frac{|\Gamma - \mathcal{F} \ \Gamma_\mathrm{sl}|}{|\Gamma_\mathrm{sl}|} = |1 - \mathcal{F}|.
\end{equation}
The following sections will introduce and compare three such methods. Also in an idealized scenario,
the correction factor should approach one as the mesh is refined (i.e., as the element size $h \to
0$), indicating that the overestimation effect vanishes with finer discretizations.

\subsection{Element size-based correction method}
\label{sec:bourdin_correction_method}
This correction method is derived from the one-dimensional analytical solution. As detailed in Section~\ref{sec:strain_localization}, the presence of strain localization causes the computed total crack surface energy, $\Gamma_{\text{sl}}$, to be overestimated by an additive term directly proportional to the element size $h$:
\begin{equation}
   \Gamma_{\text{sl}} = \Gamma + \frac{h}{2l},
\end{equation}
where $\Gamma$ is the ideal energy without localization effects. The correction factor is designed to scale the simulated energy back to the theoretical value. Based on this, the factor can be defined as:
\begin{equation}
   \mathcal{F}_\mathrm{elem} = \frac{\Gamma_{\text{sl}}}{\Gamma} = \frac{\Gamma + \frac{h}{2l}}{\Gamma} = 1 + \frac{1}{\Gamma} \frac{h}{2l}.
   \label{eq:element_size_correction_factor}
\end{equation}
A key limitation of this formulation is its dependence on $\Gamma$. In a phase-field simulation, the computed quantity is $\Gamma_{\text{sl}}$, which incorporates the strain localization effect. The objective is to correct this value to recover $\Gamma$, the physical crack area free from numerical artifacts. However, since $\Gamma$ is unknown, it cannot be used directly. To proceed, it is standard to assume the idealized scenario of a one-dimensional crack in an infinite domain (or where boundary effects are negligible, i.e., $l/a \to 0$). In this limit, the theoretical energy $\Gamma$ converges to 1, as shown in Figure~\ref{fig:CSDF_phase_field_at2_energy_l_vs_energy} and Eq.~\eqref{eq:total_energy_at2}. By adopting this assumption, the correction factor simplifies to:
\begin{equation}
   \mathcal{F}_\mathrm{Bourdin} = 1 + \frac{h}{2l}.
   \label{eq:bourdin_correction}
\end{equation}
This expression is the well-known Bourdin correction, a foundational method for mitigating crack area overestimation in phase-field models~\cite{phase_field_Bourdin2008}. It directly addresses the numerical artifact of strain localization by introducing a correction factor that is constant throughout the simulation, as it depends only on the mesh size $h$ and the length scale parameter~$l$.

The effectiveness of this method is highest when the mesh is regular in the damaged regions. As the mesh is refined ($h/l \to 0$), the correction factor approaches 1, indicating that the overestimation effect diminishes. However, a key limitation is that its derivation assumes $\Gamma \approx 1$. This assumption is only valid when the boundaries are sufficiently far compared to the length scale (i.e., $a/l \to \infty$).

This method's effectiveness is highest for straight cracks propagating perpendicular to mesh edges, a scenario that mirrors the idealized one-dimensional case from which it is derived. Furthermore, the factor does not account for boundary effects which can influence the energy term $\Gamma$. Despite these limitations, its simplicity has made it a widely adopted choice for correcting phase-field results.
It is important to note that for simulations employing symmetry boundary conditions, where only half of the crack is modeled, the theoretical energy $\Gamma$ in Equation~\eqref{eq:element_size_correction_factor} is effectively halved (i.e., $\Gamma \approx 0.5$). This implies that the Bourdin correction factor for symmetric cases becomes:
\begin{equation}
   \mathcal{F}_\mathrm{Bourdin, sym} = 1 + \frac{h}{l}.
   \label{eq:bourdin_correction_sym}
\end{equation}
This result is consistent with the correction presented in \cite{phase_field_effective_Gc_factor_2}, where the factor 2 explanation arises because the strain localization occurs over a full element width $h$, but this contribution is scaled relative to the energy of only half the crack.

\subsection{Double gradient correction method (DGCM)}
\label{sec:double_gradient_correction_method}
The Double Gradient Correction Method (DGCM) is founded on the principle that while strain localization artifacts significantly overestimate the phase-field energy term ($\Gamma_\phi$), the gradient energy term ($\Gamma_{\nabla\phi}$) remains largely unperturbed. As discussed in Section~\ref{sec:strain_localization}, strain localization causes the phase-field profile to artificially saturate at $\phi=1$ within an element. Consequently, the gradient $\nabla\phi$ becomes zero in that region, but the overall integral of the gradient term is less affected than the phase-field term.

This disparity is central. As shown in Section~\ref{sec:equipartition} and illustrated by the one-dimensional solution in Section~\ref{sec:one_dimensional_analytical_solution}, when the length scale $l$ is sufficiently small relative to the domain, the energy contributions of the phase-field and its gradient become equal: $\Gamma_\phi = \Gamma_{\nabla\phi}$. This equipartition is a fundamental property of the regularized model in the sharp-crack limit.

Since the gradient energy $\Gamma_{\nabla\phi}$ provides a more robust measure, we propose that the true physical crack area can be accurately approximated by doubling this term:
\begin{equation}
   \Gamma \approx 2 \Gamma_{\nabla\phi} = 1.
\end{equation}

This approximation effectively replaces the overestimated phase-field energy with its more reliable gradient-based counterpart.
Based on this principle, a dynamic correction factor is defined by comparing the standard simulated area, $\Gamma_\mathrm{sl} = \Gamma_{\mathrm{sl},\phi} + \Gamma_{\mathrm{sl},\nabla\phi}$, which incorporates the strain localization energy, with the proposed approximation. Recognizing that the gradient energy term is unaffected by strain localization (i.e., $\Gamma_{\mathrm{sl},\nabla\phi} = \Gamma_{\nabla\phi}$), the factor is given by:
\begin{equation}
   \mathcal{F}_\mathrm{DGCM} = \frac{\Gamma_\mathrm{sl}}{\Gamma} = \frac{\Gamma_{\mathrm{sl},\phi} + \Gamma_{\nabla\phi}}{2\Gamma_{\nabla\phi}} = \frac{1}{2} \left(1 + \frac{\Gamma_{\mathrm{sl},\phi}}{\Gamma_{\nabla\phi}}\right).
\end{equation}
This approach can be directly applied to phase-field fracture problems, where the parameters in the
correction factor are computed as integrals over the domain, obtained from the finite element
solution at each simulation step. Since the gradient term is not affected by strain localization,
whereas the phase-field term is, the correction factor ---being a function of these known
quantities--- enables the correction of the overestimation effect, given as follows:
\begin{equation}
   \mathcal{F}_\mathrm{DGCM} = \frac{1}{2} \left(1 + \frac{\int_\Omega \phi^2 \,\mathrm{d}\Omega}{l^2 \int_\Omega |\nabla \phi|^2 \,\mathrm{d}\Omega}\right).
\end{equation}

The proposed DGCM correction offers several key advantages. Its definition is independent of the mesh size ($h$) and dynamically adapts to the evolving phase-field, making it robust for unstructured meshes and complex crack paths. Crucially, its definition applies to any dimensionality (1D, 2D, 3D) without modification. In contrast, the Bourdin correction (Section~\ref{sec:bourdin_correction_method}) is derived from a one-dimensional assumption and then extended to higher dimensions.

\subsection{Skeleton correction method}
\label{sec:skeleton_correction_method}
An alternative corrective approach is the skeletonization or thresholding technique, as presented by
the authors in \cite{phase_field_Castillon2025} and based on the work of
\cite{phase_field_skeleton}. In this method, the crack area is measured directly from the finite
element mesh by identifying regions where the phase-field variable exceeds a specific threshold
(typically $\phi > 0.9$ or $0.95$). A skeletonization algorithm then extracts the central path of
the crack, representing it as a sequence of points. To improve accuracy ---particularly for cracks
propagating diagonally or exhibiting complex geometries--- a spline is fitted through these points, and the crack length is calculated from the resulting curve. This procedure accounts for sub-element crack propagation and mitigates overestimation caused by mesh discretization. Unlike the element size-based correction, the correction factor in this method is dynamic, varying throughout the simulation, and is computed at each step as:
\begin{equation}
   \mathcal{F}_\mathrm{skeleton} = \frac{\Gamma_\mathrm{sl}}{\Gamma_\mathrm{measured}},
\end{equation}
where $\Gamma_\mathrm{measured}$ is the crack area obtained through the thresholding and skeletonization procedure.

\subsection{Summary of correction methods}

This section has reviewed three methods for correcting crack area overestimation in phase-field modeling: the element size-based correction, the Double Gradient Correction Method (DGCM), and the skeletonization technique. Table~\ref{tab:crack_area_corrections} summarizes the formulas for the correction factor, $\mathcal{F}$, for each method.

\begin{table}[h!]
   \centering
   \caption{Summary of correction factors for crack area overestimation.}
   \label{tab:crack_area_corrections}
   \begin{tabular}{@{}llc@{}}
      \toprule
      \textbf{Method} & \textbf{Correction Factor ($\mathcal{F}$)} & \textbf{Type} \\
      \midrule
      Bourdin & $\mathcal{F}_\mathrm{Bourdin} = 1 + \dfrac{h}{2l}$ & Constant \\
      \addlinespace
      DGCM & $\mathcal{F}_\mathrm{DGCM} = \dfrac{1}{2} \left(1 + \dfrac{\int_\Omega \phi^2 \,\mathrm{d}\Omega}{l^2 \int_\Omega |\nabla \phi|^2 \,\mathrm{d}\Omega}\right)$ & Dynamic \\
      \addlinespace
      Skeletonization & $\mathcal{F}_\mathrm{skeleton} = \dfrac{\Gamma_\mathrm{sl}}{\Gamma_\mathrm{measured}}$ & Dynamic \\
      \bottomrule
   \end{tabular}
\end{table}

It should be noted that the application of the Bourdin correction requires that $\Gamma=1$. Similarly, the proposed method DGCM requires that the equipartition of the phase-field and gradient energies holds; thus, the accuracy of this method hinges on the validity of this assumption. These two conditions are equivalent: if energy equipartition holds (i.e., the two energy components are equal), the condition $\Gamma=1$ is satisfied, and \emph{vice versa}. The relative error associated with the application of this correction factor for the one dimensional case is given by:
\begin{equation}
   \text{Relative Error} = \frac{|2 \Gamma_{\nabla\phi} - \Gamma|}{\Gamma} = \frac{a}{l} \frac{1-\tanh^2(a/l)}{\tanh(a/l)}.
   \label{eq:relative_error_DGCM}
\end{equation}
This error reduces rapidly as the $a/l$ ratio increases; for instance, at $a/l = 5$, the error is less than $0.1\%$. This error expression is specific to the AT2 model, which has infinite support. For models with finite support (like AT1), the error is exactly zero provided the phase-field has sufficient space to develop without boundary interference.

Here, the Bourdin correction factor is a constant that depends on the mesh size and length scale. Because this correction factor is constant, it must be noted that it can be applied \textit{a priori} or \textit{a posteriori}, as the factor acts as a scaling parameter for the physical quantities. This fact can be easily analyzed in the dimensionless case or through the relationship between two problems with different critical energy release rates, as presented in \cite{phase_field_Castillon2025}. In the case of the DGCM and skeletonization methods, the correction factor is dynamically obtained at each simulation step, and it is applied \textit{a posteriori} to the simulated results.

Regarding the finite element discretization, Section~\ref{sec:fem_error_one_dimensional_problem} analyzes the energy error for the half-bar model. It should be noted that the regularization of the phase-field problem for the equivalent case where $\phi=1$ in the middle of the element has not been explicitly addressed. However, due to the strain localization effect in phase-field fracture simulations—which causes $\phi=1$ across entire elements—the discretization error is directly related to the symmetric problem analyzed in that section.

Once determined, the correction factor $\mathcal{F}$ is applied to the raw simulation results (denoted by the subscript 'sl'), which include the effects of strain localization, to recover the physical quantities. This work adopts Scheme I as the main approach, as presented in Table~\ref{tab:ch_f_factor_corrected_alternatives}.

\section{Examples}
\label{sec:examples}
To validate the proposed correction method for crack area overestimation, this section presents two benchmark examples to analyze the performance of the DGCM method and compare it against established approaches, specifically the Bourdin correction and the skeletonization technique. For this purpose, a center-cracked tension specimen and a modified compact tension specimen with internal voids are considered.
\subsection{The center cracked test specimen}
\label{example:center_cracked_test_specimen}

First, the proposed correction method is analyzed and compared with the Bourdin correction. The convergence of key quantities, such as crack length and peak force, is examined as a function of element size and the length scale parameter. Subsequently, the method's performance is evaluated in a three-dimensional context to demonstrate its applicability to 3D problems. For this purpose, the center-cracked tension specimen under force-controlled loading, as detailed in previous work by the authors~\cite{phase_field_Castillon2025}, is considered. This test case is ideal for validation, as the crack path is known a priori and analytical reference solutions from Linear Elastic Fracture Mechanics (LEFM) are available.

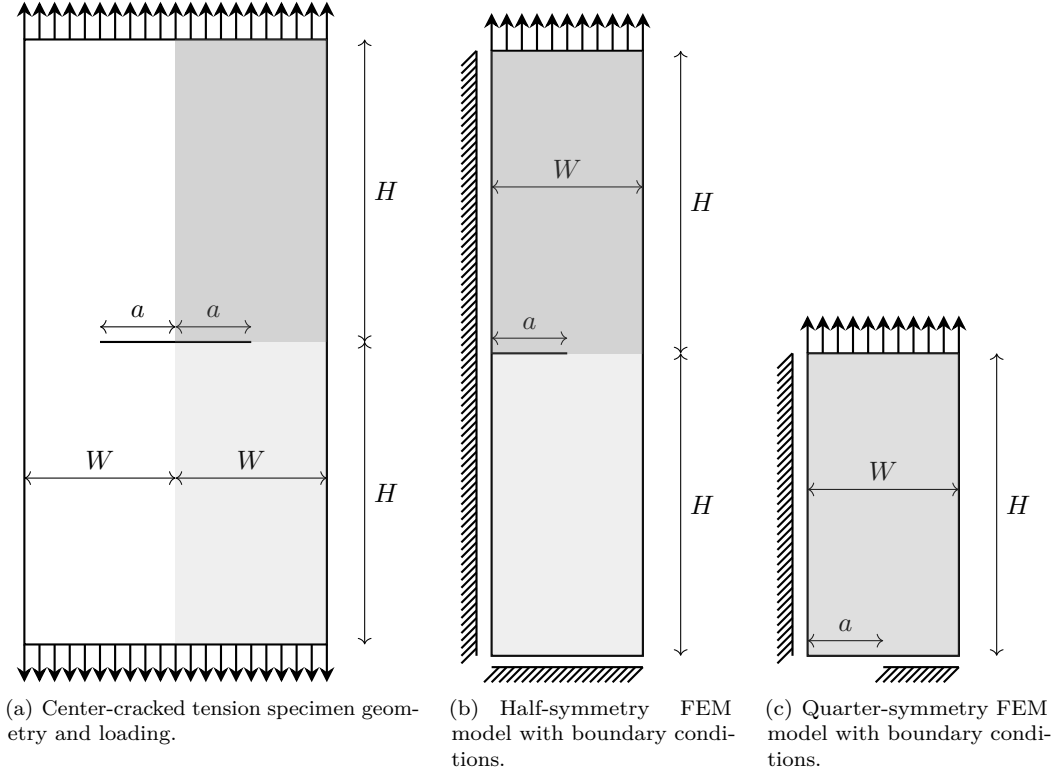
\begin{figure}[h!]
   \centering
   \subfigure[Center-cracked tension specimen geometry and loading.]{
   \begin{tikzpicture}[scale=1.0]
      \draw[thick] (0,0) rectangle (4,8);   
       \draw[thick] (1,4) -- ++(2,0); 
       \draw[<->] (1, 4.2) -- (2.0, 4.2) node[midway, above]{$a$};
       \draw[<->] (2, 4.2) -- (3, 4.2) node[midway, above]{$a$};
       \draw[<->] (0, 2.2) -- (2, 2.2) node[midway, above]{$W$};
       \draw[<->] (2, 2.2) -- (4, 2.2) node[midway, above]{$W$};
       \draw[<->] (4.5,0) -- (4.5,4) node[midway, right]{$H$};
       \draw[<->] (4.5,4) -- (4.5,8) node[midway, right]{$H$};
       \foreach \x in {0.0, 0.2, ..., 4.0}
           {
           \draw[-{Stealth[length=2mm, width=2mm]}, thick] (\x,8) -- (\x,8.5);
           }
       \foreach \x in {0.0, 0.2, ..., 4.0}
           {
           \draw[-{Stealth[length=2mm, width=2mm]}, thick] (\x,0) -- (\x,-0.5);
           }
      \fill[gray, opacity=0.25] (2,4) rectangle (4,8);
      \fill[gray, opacity=0.125] (2,0) rectangle (4,8);
   \end{tikzpicture}
   \label{fig:lefm_center_cracked_specimen}
   }
   \quad 
   \subfigure[Half-symmetry FEM model with boundary conditions.]{

      \begin{tikzpicture}[scale=1.0]
         \draw[thick] (2,0) rectangle (4,8);
       \draw[thick] (2,4) -- ++(1,0); 
       \draw[<->] (2, 4.2) -- (3, 4.2) node[midway, above]{$a$};

         \draw[<->] (2, 6.2) -- (4, 6.2) node[midway, above] {$W$};
        \draw[<->] (4.5,0) -- (4.5,4) node[midway, right]{$H$};
        \draw[<->] (4.5,4) -- (4.5,8) node[midway, right]{$H$};
         \foreach \x in {2.0, 2.2, ..., 4.0}
         {
             \draw[-{Stealth[length=2mm, width=2mm]}, thick] (\x,8) -- (\x,8.5);
         }
         \draw[thick] (4.0,-0.15) -- (2.0,-0.15);
         \foreach \x in {2.1, 2.2, ..., 4.0}
         {
             \draw[thick] (\x,-0.15) -- ++(-0.2,-0.2); 
         }
         \draw[thick] (1.8,0) -- (1.8,8);
         \foreach \y in {0.1, 0.2,..., 8} {
            \draw[thick] (1.8,\y) -- ++(-0.2,-0.2); 
         }
      \fill[gray, opacity=0.25] (2,4) rectangle (4,8);
      \fill[gray, opacity=0.125] (2,0) rectangle (4,8);
      \end{tikzpicture}
      \label{fig:fem_half_center_cracked_specimen}
   }
   \quad 
   \subfigure[Quarter-symmetry FEM model with boundary conditions.]{
      \begin{tikzpicture}[scale=1.0]
         \draw[thick] (2,4) rectangle (4,8);
         \draw[<->] (2, 4.2) -- (3, 4.2) node[midway, above] {$a$};
         \draw[<->] (2, 6.2) -- (4, 6.2) node[midway, above] {$W$};
         \draw[<->] (4.5,4) -- (4.5,8) node[midway, right] {$H$};
         \foreach \x in {2.0, 2.2, ..., 4.0}
         {
             \draw[-{Stealth[length=2mm, width=2mm]}, thick] (\x,8) -- (\x,8.5);
         }
         \draw[thick] (4.0,3.85) -- (3.0,3.85);
         \foreach \x in {3.1, 3.2, ..., 4.0}
         {
             \draw[thick] (\x,3.85) -- ++(-0.2,-0.2); 
         }
         \draw[thick] (1.8,4) -- (1.8,8);
         \foreach \y in {4.1, 4.2,..., 8} {
            \draw[thick] (1.8,\y) -- ++(-0.2,-0.2); 
         }
         \fill[gray, opacity=0.25] (2,4) rectangle (4,8);

      \end{tikzpicture}
      \label{fig:fem_quarter_center_cracked_specimen}
   }
   \caption{Schematic of the center-cracked tension test specimen. (a) Full geometry, showing dimensions and loading configuration. (b) Half-symmetry finite element model. (c) Quarter-symmetry finite element model.}
   \label{fig:center_cracked_specimen_fig}
\end{figure}

The specimen geometry and loading configuration are shown in Figure~\ref{fig:lefm_center_cracked_specimen}. It consists of a rectangular plate with a centrally located crack, subjected to uniform tensile loading. The plate has a total width of $2W = 2.0$~mm, a total height of $2H = 6.0$~mm, and an initial crack of length $2a$, with $a=0.5$~mm. The material properties are $G_c = 0.0027$~kN/mm, $E = 210$~kN/mm$^2$, and $\nu = 0.3$.

All simulations exploit the specimen's symmetry. Figure~\ref{fig:fem_half_center_cracked_specimen} shows the half-symmetry FEM model, which is used for the three-dimensional analysis in the following sections. Figure~\ref{fig:fem_quarter_center_cracked_specimen} depicts the quarter-symmetry model, which first considered under plane strain conditions, with a specimen thickness of $B = 1$~mm. In this case the left edge is constrained in the horizontal direction, while the bottom edge is fixed vertically in the region where the initial crack is absent. The top edge is subjected to an upward traction force. The energy-controlled algorithm is employed, with the traction force direction set as $\boldsymbol{t} = (0, 1/A_{\text{surface}})$~kN, where $A_\text{surface}$ is the area of the top surface where the traction is applied. The constants $c_1 = 1$~kN/mm and $c_2 = 1.0$ are used in the energy-controlled algorithm.

In this specimen, the crack propagates along a known straight horizontal path from the initial notch. This geometry allows for a straightforward characterization of boundary effects using the ratio $H/l$, where $H$ is the distance to the top boundary. With a length scale of $l=0.1$~mm and a half-height of $H=3.0$~mm, this ratio is $H/l = 30$. Since $\tanh(30) \approx 1$, the influence of the boundary is negligible. Additionally, the presence of a single, isolated crack eliminates interactions with other fracture zones. Under these conditions, the assumptions underlying both the Bourdin correction and the proposed DGCM are fully satisfied, and the theoretical relative error for the DGCM (Eq.~\eqref{eq:relative_error_DGCM}) effectively vanishes. It is worth noting that while boundary effects may be slightly more pronounced during the initial nucleation phase, they become insignificant once the crack is fully developed and propagating parallel to the distant top and bottom boundaries.

This analysis investigates the influence of the length scale parameter $l$ and the mesh size $h$. A critical aspect of this study is the ratio $l/h$, which can be realized through various combinations of these parameters. To systematically explore this relationship and facilitate comparison, we introduce the scaling factors $\alpha$ and $\theta$, defining the ratio as:
\begin{equation}
   \frac{l}{h} = \frac{\alpha l_0}{\theta h_0}
\end{equation}
where $l_0=0.0125$~mm and $h_0=0.005$~mm represent the base values for the length scale and mesh size, respectively, corresponding to Simulation 1. The specific parameter sets used for this analysis, along with all performed simulations, are detailed in Table~\ref{tab:simulations_center_cracked_specimen}. Note that all $l/h$ ratios are chosen to be greater than $2$, ensuring that the phase-field profile is sufficiently resolved in accordance with the guidelines proposed by Miehe et al.~\cite{phase_field_Miehe2010}.

\begin{table}[h!]
   \centering
   \small
   \renewcommand{\arraystretch}{1.2}
   \setlength{\tabcolsep}{8pt}
   \begin{tabular}{c c c c c c}
       \toprule
       \# & $\alpha$ & $\theta$ & Length scale $l$ (mm) & Mesh size $h$ (mm) & $l/h$ \\
       \midrule
       1  & 1.0   & 1.0    & 0.012500   & 0.005000     & 2.5 \\
       2  & 0.2   & 0.2    & 0.002500   & 0.001000     & 2.5 \\
       3  & 0.1   & 0.1    & 0.001250   & 0.000500     & 2.5 \\
       4  & 0.05  & 0.05    & 0.000625   & 0.000250    & 2.5 \\
       \addlinespace
       5 & 1.0   & 0.6250    & 0.012500   & 0.003125    & 4.0 \\
       6 & 0.2   & 0.125    & 0.002500   & 0.000625     & 4.0 \\
       7 & 0.1   & 0.0625   & 0.001250   & 0.0003125    & 4.0 \\
       8 & 0.05   & 0.03125   & 0.000625   & 0.00015625   & 4.0 \\
       \addlinespace
       9   & 0.20000   & 0.1666  & 0.002500   & 0.000833     & 3.0 \\
       10  & 0.20000   & 0.1     & 0.002500   & 0.000500     & 5.0 \\
       11  & 0.20000   & 0.0833  & 0.002500   & 0.0004166    & 6.0 \\

       \bottomrule
   \end{tabular}
   \caption{Phase-field simulation parameters for the center-cracked tension specimen. Columns $\alpha$ and $\theta$ list the scaling factors used to obtain $l$ and $h$ from base values.}
   \label{tab:simulations_center_cracked_specimen}
\end{table}

\begin{figure}[h!]
   \centering
   \subfigure[Force-displacement response.]{
      \includegraphics[width=0.45\textwidth]{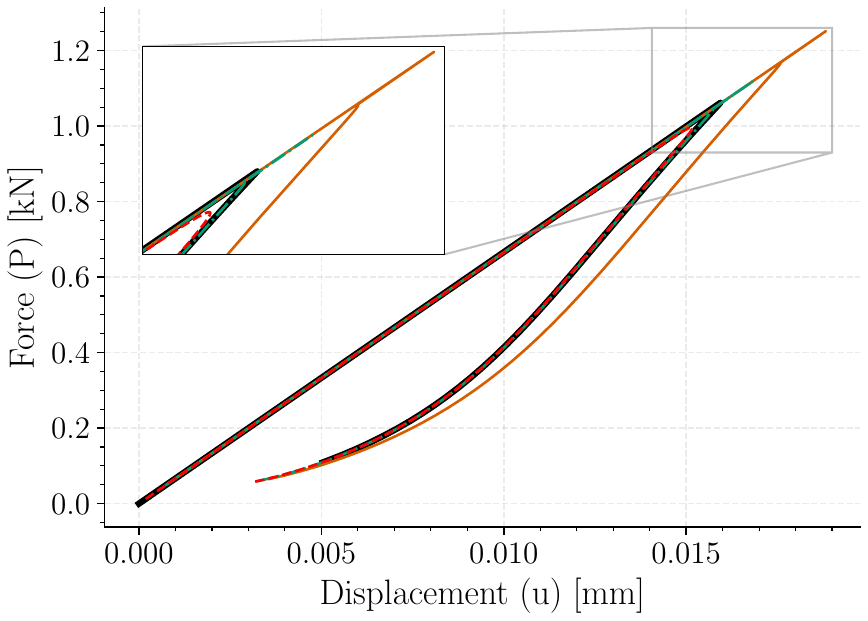}
      \label{fig:central_cracked_force_vs_displacement}
   }
   \hfill
   \subfigure[Stiffness vs. crack length.]{
      \includegraphics[width=0.45\textwidth]{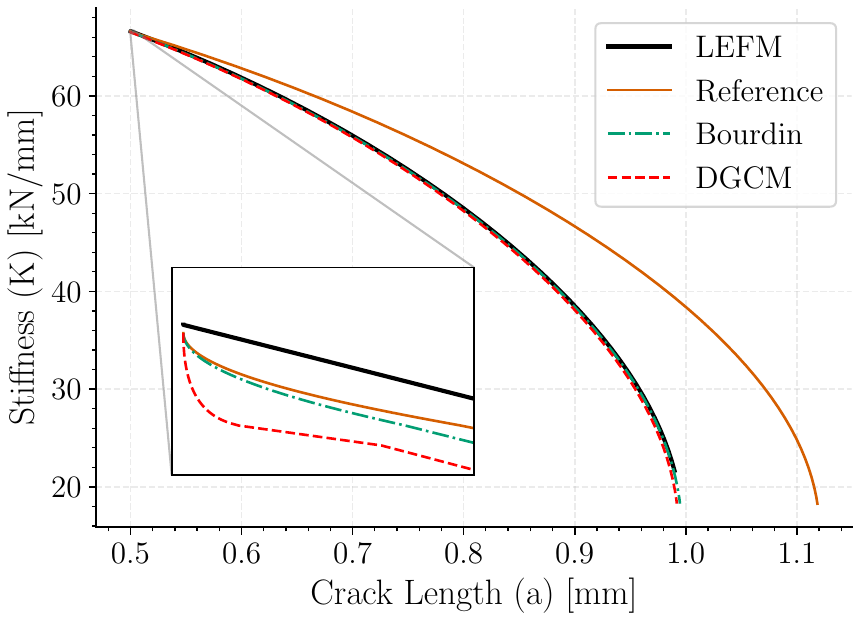}
      \label{fig:central_cracked_gamma_vs_critical_force_correction}
   }
   \caption{Comparison of numerical results with the analytical LEFM solution for the center-cracked specimen (Simulation 8). (a) Force-displacement response. (b) Stiffness degradation as a function of crack length.}
   \label{fig:compare_central_cracked_elasticity_lefm_gamma}
\end{figure}

Figure~\ref{fig:compare_central_cracked_elasticity_lefm_gamma} presents the results of Simulation 8, comparing the reference solution, the solutions obtained after applying the post-processing correction methods, and the analytical LEFM solution. Figure~\ref{fig:central_cracked_force_vs_displacement} shows the force-displacement response. As observed, the results obtained with either the Bourdin correction or the proposed DGCM are in good agreement with the LEFM solution, whereas the uncorrected results deviate significantly. It is also important to note the presence of a peak force overshoot in the reference solution, a phenomenon reported by other authors~\cite{phase_field_snap_pedro, phase_field_snap_Ritukesh, phase_field_snap_Zambrano}. This overshoot persists with the Bourdin correction, as it acts as a constant scaling factor. In contrast, the proposed DGCM effectively eliminates this artifact, acting as a regularization mechanism. Figure~\ref{fig:central_cracked_gamma_vs_critical_force_correction} displays the stiffness as a function of crack length. Again, the corrected solutions match the LEFM prediction well. A zoomed view of the initial crack propagation phase reveals that, in all cases, a loss of stiffness occurs due to damage accumulation during loading.

To rigorously analyze the convergence of the correction factors, two key quantities are examined: the peak force value and the crack length corresponding to a fixed stiffness. This analysis is conducted in two stages. First, the influence of the length scale parameter $l$ is investigated by decreasing $l$ while maintaining a constant ratio $l/h$. Second, the effect of mesh refinement is isolated by varying the element size $h$ while keeping the length scale parameter $l$ constant.

\subsubsection{Convergence analysis with respect to length scale $l$ at constant mesh resolution $l/h$}
\begin{figure}[h!]
   \centering
   \subfigure[Stiffness vs. applied force.]{
     \includegraphics[width=0.45\textwidth]{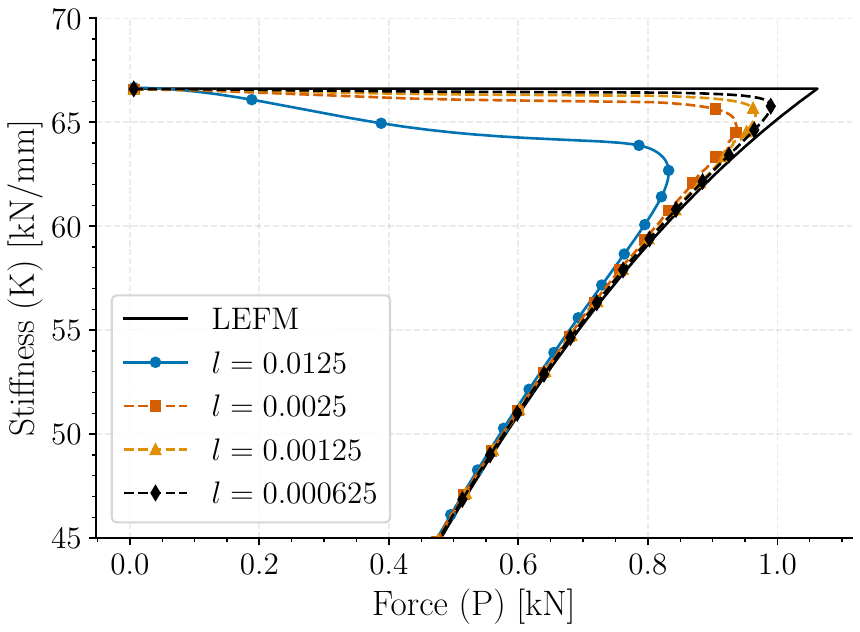}
     \label{fig:stiffness_vs_force_lh_2_5}
   }
   \hfill
   \subfigure[Correction factor evolution vs. crack length.]{
   \includegraphics[width=0.45\textwidth]{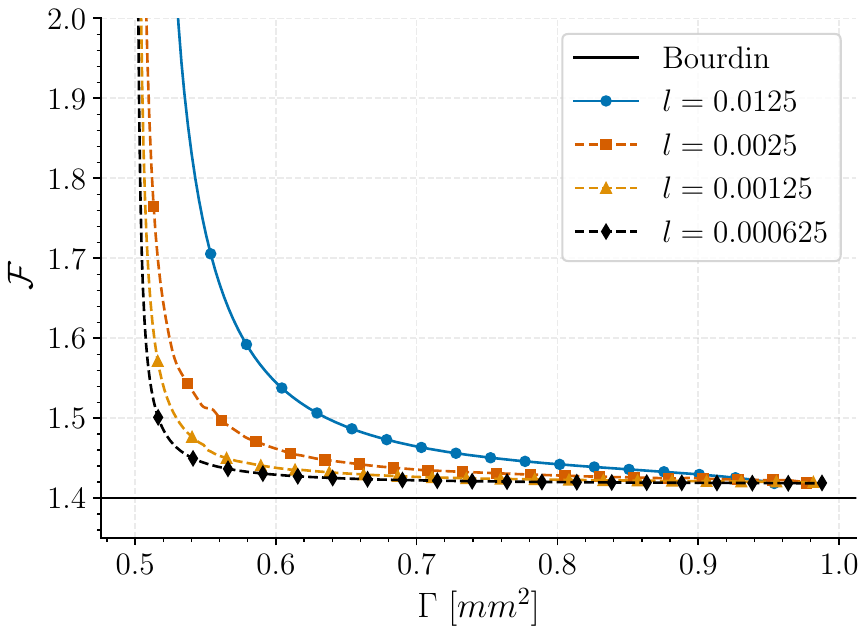}
     \label{fig:compare_correction_factor_lh}
   }
   \caption{Convergence analysis for a constant mesh resolution ratio $l/h=2.5$. (a) Stiffness degradation as a function of applied force for decreasing length scales $l$, illustrating convergence toward the LEFM solution. (b) Evolution of the DGCM correction factor with crack length compared to the constant Bourdin factor.}
   \label{fig:correction_factor0}
\end{figure}

First, the influence of the length scale parameter $l$ is analyzed. This study is conducted using a fixed mesh resolution ratio of $l/h=2.5$, ensuring that the phase-field profile remains well-resolved and that the element size scales proportionally with the length scale. The simulations corresponding to $l/h=2.5$ are detailed in rows 1-4 of Table~\ref{tab:simulations_center_cracked_specimen},

Before presenting the formal convergence analysis, we examine several key curves to provide an intuitive understanding of the model's behavior. Figure~\ref{fig:stiffness_vs_force_lh_2_5} illustrates the specimen stiffness as a function of the applied force using the proposed DGCM. As the length scale parameter decreases, the onset of damage during the initial loading phase is delayed, and the response converges toward a constant stiffness that matches the LEFM prediction. Furthermore, the graph demonstrates that as $l$ decreases, the maximum force at crack initiation converges to a stable value, closely aligning with the theoretical prediction. Notably, the proposed method eliminates the force overshoot observed in both the uncorrected reference results and the Bourdin correction.

Figure~\ref{fig:compare_correction_factor_lh} displays the DGCM correction factors as a function of the corrected crack length. Since these simulations share the same $l/h$ ratio, the Bourdin correction factor remains constant across all cases; this is represented by a solid horizontal line on the graph. As the length scale is reduced, the initial transient region of the curve—associated with crack nucleation and the peak force overshoot—becomes sharper, eventually converging to a linear trend. This indicates that in the limit as $l \to 0$, the correction factor rapidly stabilizes.

Having analyzed the general trends of the correction factors, we now examine the convergence behavior of the peak force and the crack length corresponding to a fixed stiffness value. This analysis compares the uncorrected reference solution with results obtained using the Bourdin and DGCM correction methods.

\begin{figure}[h!]
   \centering
   \subfigure[Convergence of the peak force.]{
     \includegraphics[width=0.45\textwidth]{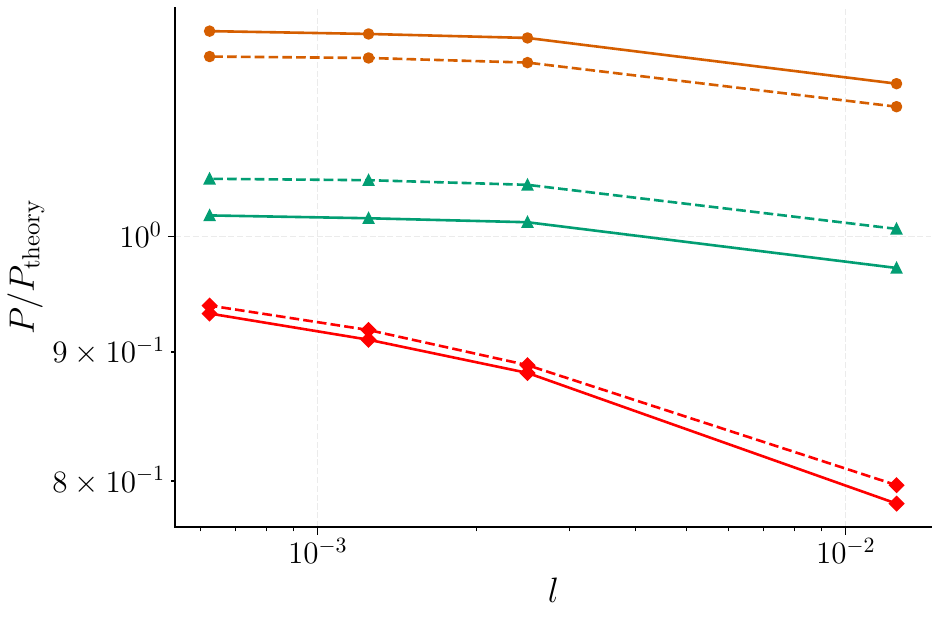}
     \label{fig:convergence_h_constant_l_max_force2}
   }
   \hfill
   \subfigure[Convergence of crack length at fixed stiffness ($K \approx 38.5$~kN/mm).]{
     \includegraphics[width=0.45\textwidth]{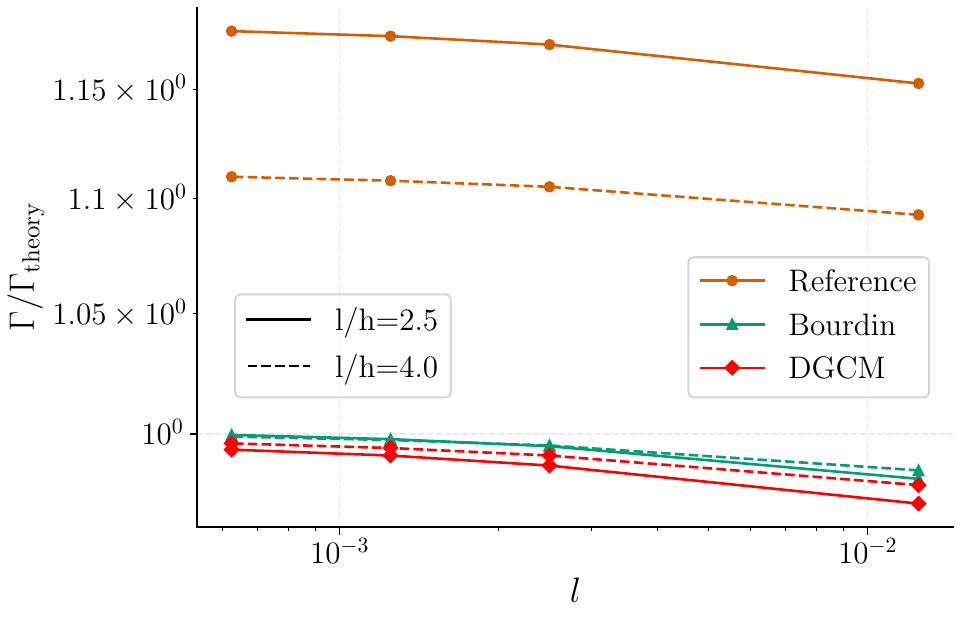}
     \label{fig:convergence_h_constant_l_gamma_at_stiffness_38_525142}
   }
   \caption{Convergence analysis with respect to the length scale parameter $l$ for constant mesh resolution ratios ($l/h=2.5$ and $l/h=4.0$). (a) Peak force convergence toward the LEFM limit. (b) Crack length convergence for a specific stiffness value.}
   \label{fig:convergence_h_constant_l2}
\end{figure}

Figure~\ref{fig:convergence_h_constant_l_max_force2} depicts the peak force as a function of the length scale parameter $l$ for the reference and corrected solutions. The uncorrected reference solution exhibits the highest force values, significantly inflated by the characteristic force overshoot. The Bourdin correction, acting as a constant scaling factor, reduces the peak force magnitude but fails to eliminate the overshoot artifact. In contrast, the proposed DGCM correction not only significantly reduces the peak force but also effectively suppresses the overshoot. As the length scale $l$ decreases, the DGCM-corrected peak force converges toward the theoretical value predicted by Linear Elastic Fracture Mechanics (LEFM). Conversely, while the other methods show a decreasing trend with smaller $l$, they consistently overestimate the theoretical peak force.

Figure~\ref{fig:convergence_h_constant_l_gamma_at_stiffness_38_525142} illustrates the convergence of the crack length for a fixed stiffness value of $38.52514$~kN/mm. The theoretical crack length corresponding to this stiffness is $0.9$~mm. Since this measurement point is well beyond the initial crack nucleation phase and the associated force overshoot, both the Bourdin and DGCM corrections yield similar results. As the length scale parameter decreases, both methods converge toward the theoretical crack length.

In addition to the primary analysis with $l/h=2.5$ (solid lines), Figures~\ref{fig:convergence_h_constant_l_max_force2} and \ref{fig:convergence_h_constant_l_gamma_at_stiffness_38_525142} also display results for a finer mesh resolution ratio of $l/h=4.0$ (dashed lines). For the peak force, the finer mesh results in higher values for both correction methods compared to the coarser mesh. However, for the uncorrected solution, the peak force decreases with mesh refinement. It is worth noting that as the ratio $h/l \to 0$, the Bourdin correction factor approaches unity, causing the corrected and uncorrected solutions to coincide. A similar convergence behavior is observed for the crack length measurements.

\begin{figure}[h!]
   \centering
   \includegraphics[width=8cm]{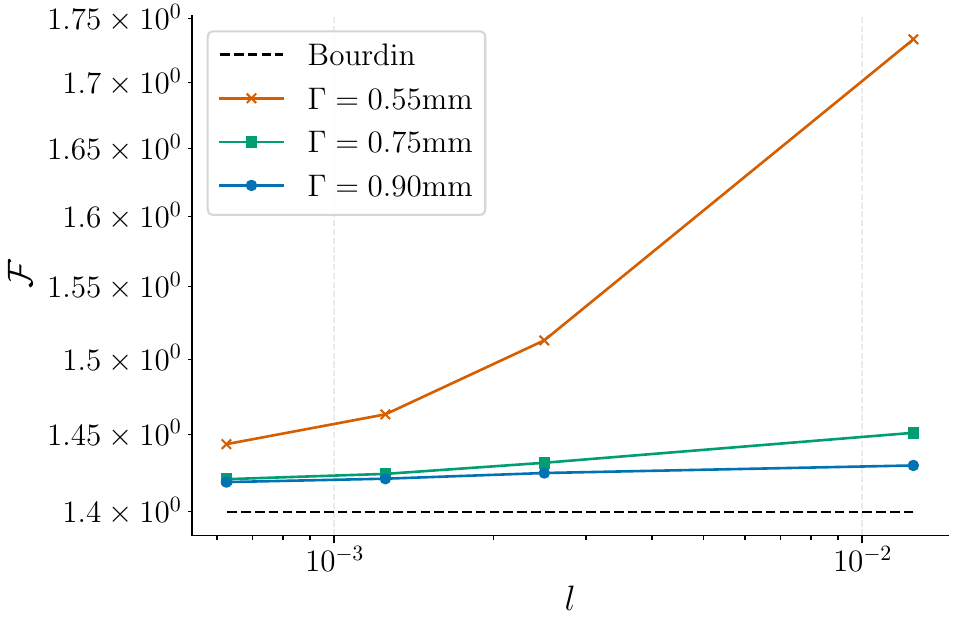}
   \caption{Convergence of the DGCM correction factor as a function of the length scale parameter $l$ for different crack lengths.}
   \label{fig:convergence_correction_factor_lh}
\end{figure}

Finally, Figure~\ref{fig:convergence_correction_factor_lh} shows the convergence of the DGCM correction factor evaluated at three distinct crack lengths: $0.55$~mm, $0.75$~mm, and $0.9$~mm. Additionally, the Bourdin correction factor is plotted; since the ratio $l/h$ is held constant in this analysis, this factor appears as a constant horizontal line. As the length scale parameter decreases, the DGCM correction factor drops rapidly, indicating that the relative overestimation of the crack area diminishes with finer length scales. Furthermore, at crack lengths far from the initial nucleation zone (e.g., $0.75$~mm and $0.9$~mm), the correction factors converge to similar values, as the influence of the initial peak force overshoot dissipates.

\subsubsection{Convergence analysis with respect to mesh size $h$ at constant length scale $l$}

In this case, the length scale parameter is fixed at $l=0.0025$~mm, while the mesh size $h$ is varied to study the influence of mesh refinement. The simulations corresponding to this analysis are detailed in rows 2, 6, 9, 10, and 11 of Table~\ref{tab:simulations_center_cracked_specimen}.

\begin{figure}[h!]
   \centering
   \subfigure[Convergence of the peak force.]{
     \includegraphics[width=0.45\textwidth]{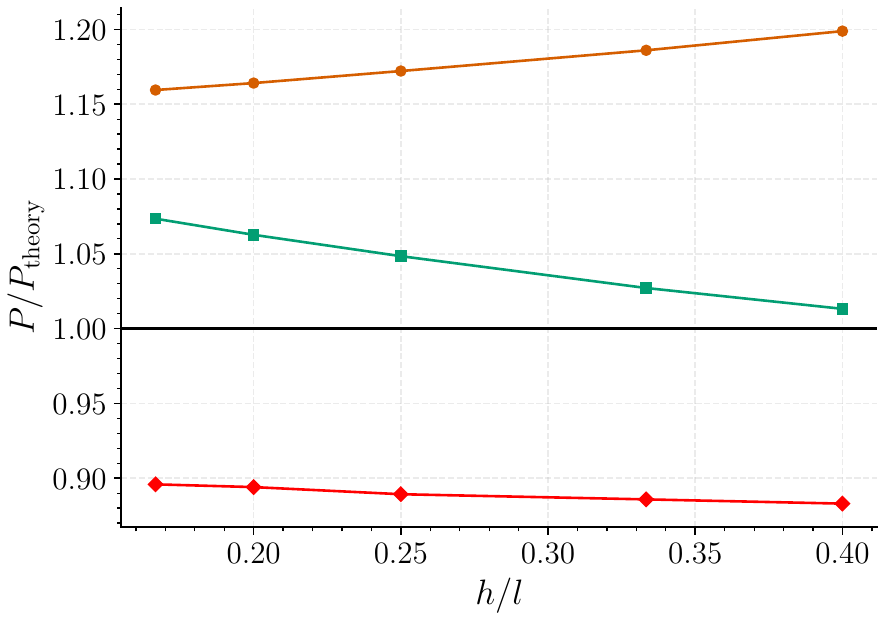}
     \label{fig:convergence_h_constant_l_max_force}
   }
   \hfill
   \subfigure[Convergence of crack length at fixed stiffness ($K \approx 38.5$~kN/mm).]{
     \includegraphics[width=0.45\textwidth]{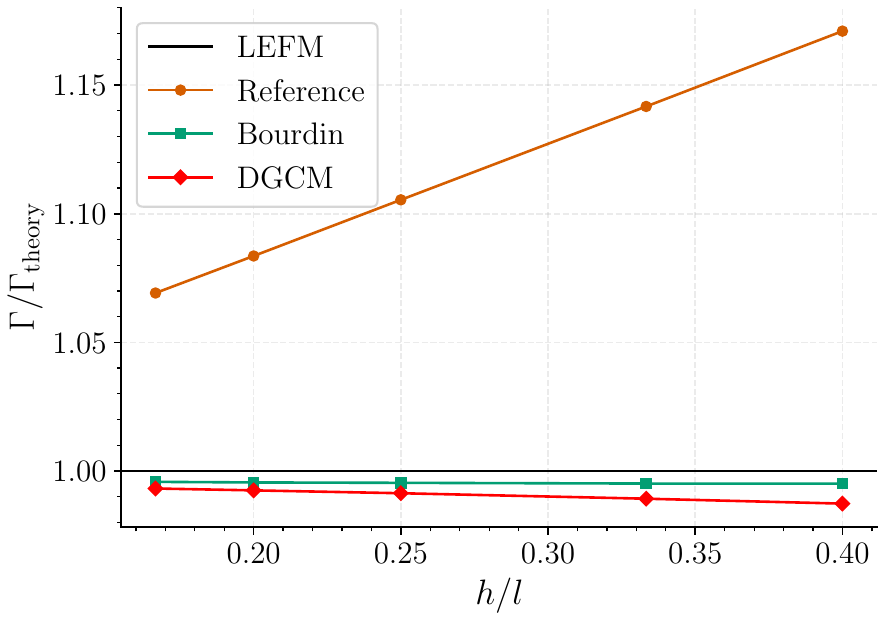}
     \label{fig:convergence_h_constant_l_gamma_at_stiffness_38_52514}
   }
   \caption{Convergence analysis with respect to mesh size $h$ for a constant length scale $l=0.0025$~mm. (a) Peak force convergence. (b) Crack length convergence for a specific stiffness value.}
   \label{fig:convergence_h_constant_l}
\end{figure}

Figure~\ref{fig:convergence_h_constant_l_max_force} shows the evolution of the peak force as a function of mesh size $h$ for a fixed length scale $l=0.0025$~mm. As the mesh is refined (i.e., $h$ decreases), the Bourdin correction factor approaches 1, causing the Bourdin-corrected solution to converge toward the uncorrected result. Notably, the uncorrected solution exhibits a decreasing trend in peak force with mesh refinement, while the Bourdin-corrected peak force increases as the mesh size becomes smaller. This indicates a tendency for the peak force to increase with decreasing mesh size when the Bourdin correction is applied. In contrast, the proposed DGCM correction consistently yields a lower and more stable peak force, effectively eliminating the artificial peak observed in standard phase-field simulations. Furthermore, the DGCM-corrected peak force demonstrates reduced sensitivity to mesh refinement, indicating improved robustness and convergence with respect to mesh size. It is also observed that, as the mesh is refined, both the uncorrected and Bourdin-corrected results converge toward a common value that lies between them, with the DGCM-corrected result closely matching this converged value.

Figure \ref{fig:convergence_h_constant_l_gamma_at_stiffness_38_52514} illustrates the convergence of the crack length for a fixed stiffness value of $38.52514$~kN/mm as the mesh is refined. Similar to the previous analysis, both the Bourdin and DGCM corrections yield comparable results for this measurement point, which is well beyond the initial crack nucleation phase. As the mesh size decreases, both correction methods converge toward the theoretical crack length of $0.9$~mm.

\begin{figure}[h!]
   \centering
   \includegraphics[width=8cm]{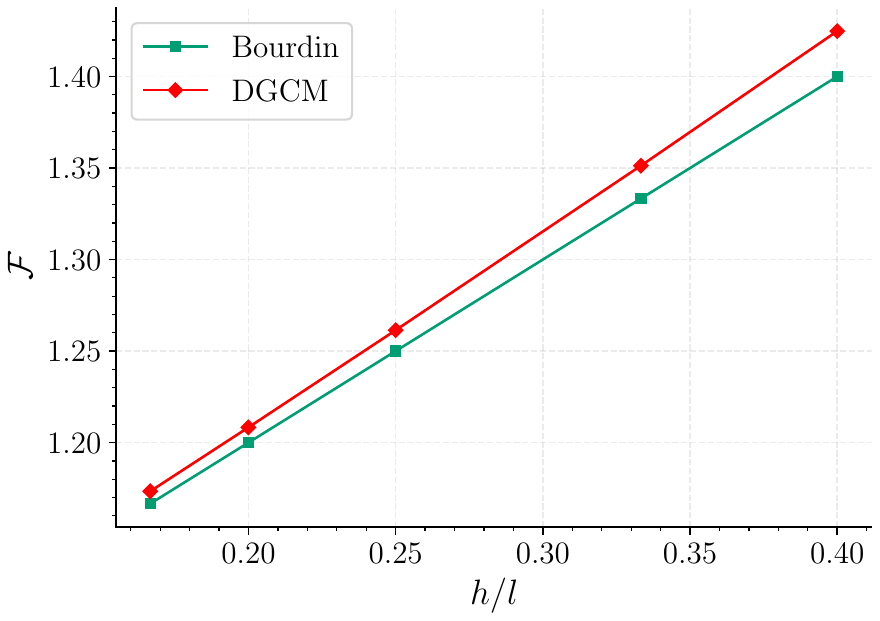}
    \caption{Convergence of the Bourdin and DGCM correction factors as a function of mesh size $h$, evaluated at a crack length of $0.9$~mm.}
   \label{fig:convergence_correction_factor_h}
\end{figure}

Figure~\ref{fig:convergence_correction_factor_h} shows the convergence of the Bourdin and DGCM correction factors as a function of mesh size $h$, evaluated at a crack length of $0.9$~mm. As the mesh is refined (i.e., as $h$ decreases), both correction factors decrease and approach unity, indicating that the overestimation effect diminishes with finer meshes.

\subsubsection{Extension to 3D}
To demonstrate the applicability and robustness of the DGCM in three-dimensional scenarios, the analysis is extended to a 3D version of the center-cracked tension specimen. The half-symmetry model shown in Figure~\ref{fig:fem_half_center_cracked_specimen}, with an assigned thickness of $B = 0.05$~mm, is employed. The boundary conditions are defined as follows: the left face is constrained horizontally to enforce symmetry, and the bottom face is constrained in the vertical and out-of-plane directions. The mesh consists of tetrahedral elements with a characteristic size of $h \approx 0.0025$~mm in the expected crack propagation zone. A length scale of $l=0.00625$~mm is used, resulting in a mesh resolution of $l/h \approx 2.5$. The simulation utilizes the non-variational energy-controlled scheme with constants $c_1 = 1.5$~kN/mm and $c_2 = 1.0$, keeping material properties consistent with the 2D analysis.

This validation is particularly significant because 3D simulations often involve complex, unstructured meshes where methods like skeletonization are challenging to implement. Although it might be possible to approximate the crack area in this specific example by applying skeletonization to one of the external faces, the proposed DGCM provides a direct, volume-based measurement without such complexities. Furthermore, the Bourdin correction is typically evaluated for one-dimensional or two-dimensional quadrilateral elements, implying a straightforward extension to hexahedral elements in 3D. In contrast, the assumption underlying the proposed method applies universally to all element types in one, two, or three dimensions without requiring additional considerations.

\begin{figure}[h!]
   \centering
   \subfigure[Initial propagation]{
      \includegraphics[width=0.3\linewidth]{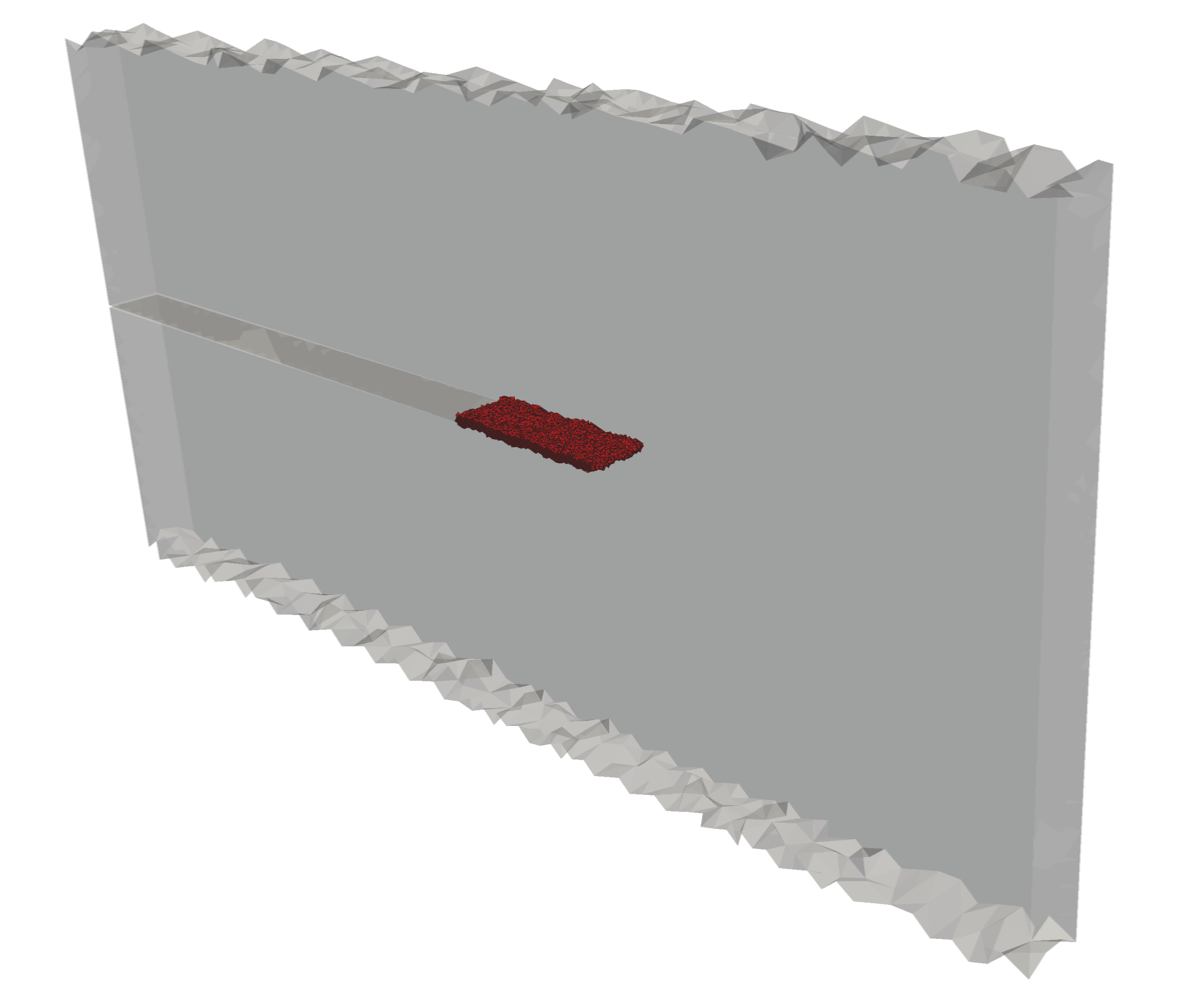}
      \label{fig:center_cracked_3d_stage1}
   }
   \hfill
   \subfigure[Mid-propagation]{
      \includegraphics[width=0.30\linewidth]{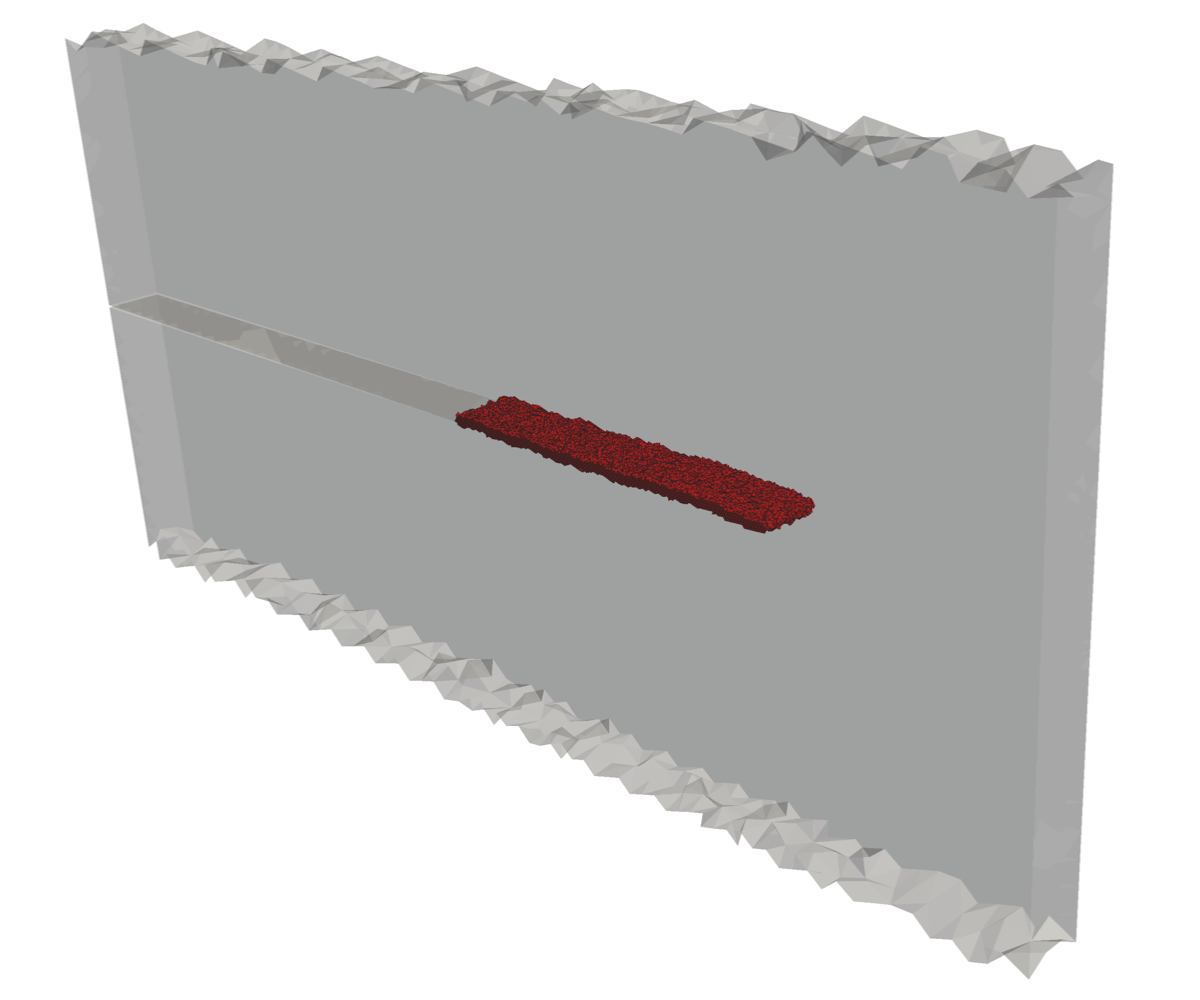}
      \label{fig:center_cracked_3d_stage2}
   }
   \hfill
   \subfigure[Final stage]{
      \includegraphics[width=0.30\linewidth]{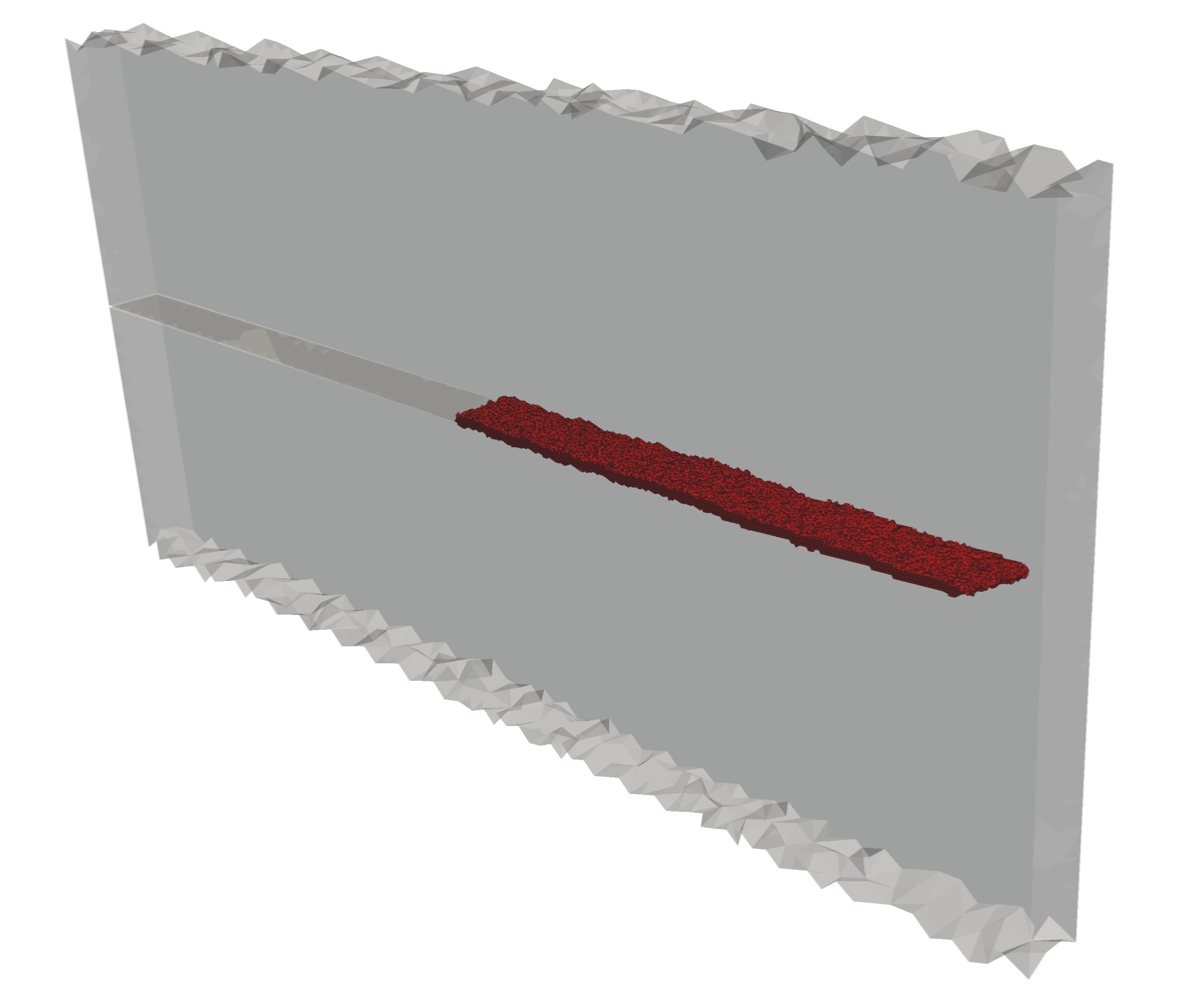}
      \label{fig:center_cracked_3d_stage3}
   }
   \caption{Crack propagation in the 3D center-cracked tension specimen. The images show the isosurface of the phase-field variable for $\phi > 0.95$ at three different stages of the simulation, illustrating the evolution of the crack front.}
   \label{fig:center_cracked_3d_crack_path}
\end{figure}

The evolution of the crack front is visualized in Figure~\ref{fig:center_cracked_3d_crack_path}, which displays the isosurfaces where the phase-field variable $\phi > 0.95$ at different stages of the simulation.

Table~\ref{tab:3d_crack_area_evolution_with_error} presents the crack area obtained using the
different correction methods at three distinct stages of crack propagation. Additionally, the
relative error of the reference, Bourdin, and DGCM methods with respect to the skeletonization
measurement is shown. As observed, the proposed DGCM yields the smallest relative error. In general, errors are larger for all methods during the initial stage, where crack nucleation and peak force effects are more pronounced. These results demonstrate the effectiveness of the DGCM in accurately estimating the crack area in a 3D context, significantly outperforming the Bourdin correction while avoiding the complexities associated with skeletonization on unstructured meshes.
\begin{table}[h!]
   \centering
   \caption{Evolution of the crack area and relative error (in parentheses) with respect to the Skeletonization method in the 3D center-cracked specimen.}
   \label{tab:3d_crack_area_evolution_with_error}
   \renewcommand{\arraystretch}{1.2}
   \begin{tabular}{@{}lccc@{}}
      \toprule
      \textbf{Correction Method} & \textbf{Initial Prop.} (mm$^2$) & \textbf{Mid Prop.} (mm$^2$) & \textbf{Final Stage} (mm$^2$) \\
      \midrule
      Skeletonization         & 0.00707 & 0.01493 & 0.02222 \\
      Reference               & 0.01035 (46.26\%) & 0.02051 (37.36\%) & 0.03057 (37.63\%) \\
      Bourdin                 & 0.00862 (21.88\%) & 0.01709 (14.47\%) & 0.02548 (14.69\%) \\
      DGCM                    & 0.00752 (6.24\%) & 0.01515 (1.45\%) & 0.02278 (2.54\%) \\
      \bottomrule
   \end{tabular}
\end{table}

\subsection{Modified compact tension test specimen}
\label{example:compact_tension_test_specimen}
This section validates the proposed correction factor by comparing its performance against the Bourdin and skeletonization methods in a complex scenario involving curvilinear crack paths. The analysis employs modified compact tension (CT) specimens with internal voids, replicating a challenging benchmark from the experimental campaign by Wagner et al. \cite{example_Wagner2019_article, example_Wagner2018_phd_thesis}.

\begin{figure}[h!]
   \centering

\resizebox{0.6\textwidth}{!}{%
\begin{tikzpicture}[scale=0.2]  
   \def\b{40}
   \def\hg{0.6*\b}         
   \def\a{0.2*\b}          
   \def\hone{0.275*\b}     
   \def\c{0.25*\b}         
   \def\D{0.25*\b}         
   \def\H{2.0}             
   
   \def\Da{0.2*\b}         
   \def\xa{0.45*\b}        
   \def\ha{0.25*\b}        
   
   \def\Db{0.1*\b}         
   \def\xb{0.75*\b}        
   \def\hb{0}              
   
   \def\Dc{0.2*\b}         
   \def\xc{0.625*\b}       
   \def\hc{-0.25*\b}       
   
   \def\f{0.06375*\b}      
   \def\j{0.08875*\b}      
   \def\k{0.04250*\b}      
   \def\l{0.015*\b}        
   \def\g{\a - \l/0.57735}   
   

   \draw[very thick, black] (-\c,-\hg) -- (\b,-\hg);
   \draw[very thick, black] ( \b,-\hg) -- (\b, \hg);
   \draw[very thick, black] (-\c, \hg) -- (\b, \hg);

   \draw[very thick, black] (-\c,\hg) -- (-\c,\H+\f);
   \draw[very thick, black] (-\c,\H-\f) -- (-\c,-\hg);
   
   \draw[very thick, black] (0,\hone) circle (\D/2);
   \draw[thick, black] ({-0.5}, \hone) -- ({0.5}, \hone);
   \draw[thick, black] (0, {\hone-0.5}) -- (0, {\hone+0.5});
   
   \draw[very thick, black] (0,-\hone) circle (\D/2);
   \draw[thick, black] ({-0.5}, {-\hone}) -- ({0.5}, {-\hone});
   \draw[thick, black] (0, {-\hone-0.5}) -- (0, {-\hone+0.5});
   
   \draw[very thick, black] (\xa,\ha) circle (\Da/2);
   \draw[thick, black] ({\xa-0.5}, \ha) -- ({\xa+0.5}, \ha) node[right] {a};
   \draw[thick, black] (\xa, {\ha-0.5}) -- (\xa, {\ha+0.5});
   \draw[very thick, black] (\xb,\hb) circle (\Db/2) node[left] {b};
   \draw[thick, black] ({\xb-0.5}, \hb) -- ({\xb+0.5}, \hb);
   \draw[thick, black] (\xb, {\hb-0.5}) -- (\xb, {\hb+0.5});

   \draw[very thick, black] (\xc,\hc) circle (\Dc/2);
   \draw[thick, black] ({\xc-0.5}, \hc) -- ({\xc+0.5}, \hc) node[right] {c};
   \draw[thick, black] (\xc, {\hc-0.5}) -- (\xc, {\hc+0.5});
   
   \draw[thick, black] (-\c,\H+\f) -- (-\c+\k,\H+\j);
   \draw[thick, black] (-\c+\k,\H+\j) -- (-\c+\k,\H+\l);
   \draw[thick, black] (-\c+\k,\H+\l) -- (\g,\H+\l);
   \draw[thick, black] (\g,\H+\l) -- (\a,\H);
   \draw[thick, black] (\a,\H) -- (\g,\H-\l);
   \draw[thick, black] (\g,\H-\l) -- (-\c+\k,\H-\l);
   \draw[thick, black] (-\c+\k,\H-\l) -- (-\c+\k,\H-\j);
   \draw[thick, black] (-\c+\k,\H-\j) -- (-\c,\H-\f);

   \draw[<->, blue] (0,\H+1.5) -- (\a, \H+1.5) node[midway, above, blue] {$0.2 W$};
   \draw[<->, blue] (-\c,-\hg-3) -- (0,-\hg-3) node[midway, below, blue] {$0.25 W$};
   \draw[<->, blue] (0,-\hg-3) -- (\b,-\hg-3) node[midway, below, blue] {$W$};
   
   \draw[<->, blue] (\b+2, 0 ) -- (\b+2, \hg) node[midway, right, blue] {$0.6 W$};
   \draw[<->, blue] (\b+2, 0 ) -- (\b+2,-\hg) node[midway, right, blue] {$0.25 W$};

   \draw[<->, blue] (-\c-4, \hg) -- (-\c-4, \H) node[midway, left, blue] {$H$};
   \draw[dashed, blue] (-\c-4, \H) -- (\a, \H);
   \draw[dashed, blue] (-\c-4, \hg) -- (-\c, \hg);
   
   \draw[<-, blue] ({\D/2*cos(45)}, {\hone - \D/2*sin(45)}) -- ({\D/2*cos(45) + 0.5}, {\hone - \D/2*sin(45) + 0.5}) -- ({\D/2*cos(45) + 1.5}, {\hone - \D/2*sin(45) + 0.5}) node[right, blue] {$\diameter 0.25 W$};
   \draw[<-, blue] ({\D/2*cos(45)}, {-\hone + \D/2*sin(45)}) -- ({\D/2*cos(45) + 0.5}, {-\hone + \D/2*sin(45) + 0.5}) -- ({\D/2*cos(45) + 1.5}, {-\hone + \D/2*sin(45) + 0.5}) node[right, blue] {$\diameter 0.25 W$};

   \draw[<-, blue] ({\xa + \Da/2*cos(45)}, {\ha + \Da/2*sin(45)}) -- ({\xa + \Da/2*cos(45) + 0.5}, {\ha + \Da/2*sin(45) + 0.5}) -- ({\xa + \Da/2*cos(45) + 1.5}, {\ha + \Da/2*sin(45) + 0.5}) node[right, blue] {$\diameter 0.2 W$};

   \draw[<-, blue] ({\xb - \Db/2*cos(45)}, {\hb + \Db/2*sin(45)}) -- ({\xb - \Db/2*cos(45) - 0.5}, {\hb + \Db/2*sin(45) + 0.5}) -- ({\xb - \Db/2*cos(45) - 1.5}, {\hb + \Db/2*sin(45) + 0.5}) node[left, blue] {$\diameter 0.1 W$};

   \draw[<-, blue] ({\xc + \Dc/2*cos(45)}, {\hc + \Dc/2*sin(45)}) -- ({\xc + \Dc/2*cos(45) + 0.5}, {\hc + \Dc/2*sin(45) + 0.5}) -- ({\xc + \Dc/2*cos(45) + 1.5}, {\hc + \Dc/2*sin(45) + 0.5}) node[right, blue] {$\diameter 0.2 W$};
   
   \draw[<->, blue] (-\D/2-2.25,\hone) -- (-\D/2-2.25,\hg) node[pos=0.75, right, blue] {$0.325 W$};
   \draw[<->, blue] (-\D/2-2.25,-\hg) -- (-\D/2-2.25,-\hone) node[pos=0.25, right, blue] {$0.325 W$};
   
   \draw[dashed, blue] (0, -\hg-3) -- (0, \hg);
   \draw[dashed, blue] (-\D/2-2.25,-\hone) -- (0,-\hone);
   \draw[dashed, blue] (-\D/2-2.25, \hone) -- (0, \hone);

   \draw[<->, blue] (\xa-\Da/2-1.25,\ha) -- (\xa-\Da/2-1.25,\hg) node[midway, right, blue] {$0.35 W$};
   \draw[<->, blue] (\xa,\ha+\Da/2+1.2) -- (\b,\ha+\Da/2+1.2) node[midway, above, blue] {$0.55 W$};

   \draw[dashed, blue] (\xa,\ha+\Da/2+1.2) -- (\xa,\ha);
   \draw[dashed, blue] (\xa-\Da/2-1.25,\ha) -- (\xa,\ha);

   \draw[<->, blue] (\xb,\hb+\Db/2+1.2) -- (\b,\hb+\Db/2+1.2) node[midway, above, blue] {$0.25 W$};
   \draw[dashed, blue] (\xb,\hb+\Db/2+1.2) -- (\xb,\hb);
   \draw[dashed, blue] (\b+2, 0) -- (\xb,\hb);

   \draw[<->, blue] (\xc-\Dc/2-1.25,\hc) -- (\xc-\Dc/2-1.25,-\hg) node[midway, left, blue] {$0.35 W$};
   \draw[<->, blue] (\xc,\hc-\Dc/2-1.2) -- (\b,\hc-\Dc/2-1.2) node[midway, below, blue] {$0.375 W$};

   \draw[dashed, blue] (\xc-\Dc/2-1.25,\hc) -- (\xc,\hc);
   \draw[dashed, blue] (\xc,\hc-\Dc/2-1.2) -- (\xc,\hc);

   

    \draw[<-, blue] (-2,\H-\l) -- (-2,\H-\f) node[midway, below, xshift=-10pt, yshift=-5pt, blue] {$0.0075 W$};
   \draw[<-, blue] (-2,\H+\l) -- (-2,\H+\f);



   \draw[thin, blue, dashed] (\a,\H) -- ({\a+2.6*cos(30)},{\H+2.6*sin(30)});
   \draw[thin, blue, dashed] (\a,\H) -- ({\a+2.6*cos(30)},{\H-2.6*sin(30)});

   \draw[<->, thick, blue] ({\a+2.0*cos(30)},{\H+2.0*sin(30)}) arc[start angle=30, end angle=-30, radius=2.0] node[midway, right, blue] {$60^\circ$};

\foreach \angle in {180, 190, ..., 360}
{
    \draw[thick, red] ({0 + (\D/2)*cos(\angle)}, {-\hone + (\D/2)*sin(\angle)}) 
      -- ++({-(0.8/40)*\b*cos(\angle)}, {-(0.8/40)*\b*sin(\angle)});
}
\node[anchor=north, red] at (\D/2, -\hone - \D/2 - 0) {$u_x = u_y = 0$};

\foreach \angle in {10, 30,..., 170}
{
    \draw[->, thick, red] ({0 + (\D/2)*cos(\angle)}, {\hone + (\D/2)*sin(\angle)}) 
         -- ++({0}, {(2.4/40)*\b});
}

\end{tikzpicture}
}
\caption{Configuration of the compact tension specimen. The complete geometry is shown with all dimensions labeled in blue, relative to the specimen width $W=40.0$~mm. Boundary conditions, labeled in red, indicate the fixed constraint at the bottom loading hole and the applied forces at the top loading hole.}
\label{fig:compact_specimen_configuration}
\end{figure}
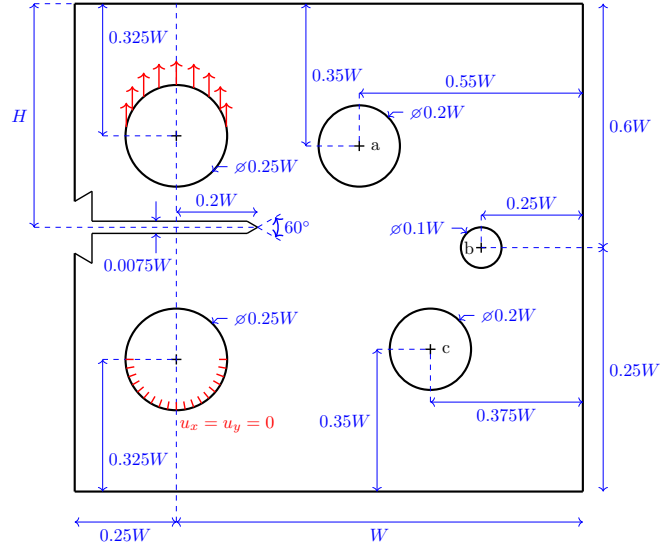

The specimen geometry, based on the ASTM E647 standard, is adapted to include additional holes and varied initial crack positions ($H$), as detailed in Figure~\ref{fig:compact_specimen_configuration}. The specimen has a width of $W=40.0$~mm, and all other dimensions are defined relative to this width. The material is a Haynes 230 superalloy with the following properties: $E=211$~kN/mm$^2$, $\nu=0.3$, and $G_c=0.073$~kN/mm. The analysis presented here follows the work of the authors in \cite{phase_field_Castillon2025}, using the same material and simulation parameters for validation.

Three configurations are analyzed: the standard CT specimen without additional holes (Specimen 1) and two modified specimens with varying initial crack positions. For Specimen 1, the initial crack is centered ($H=0.60W$), resulting in a symmetric configuration. The other two configurations are Specimen 2 ($H=0.56W$) and Specimen 4 ($H=0.64W$). To maintain consistency with the numbering in \cite{phase_field_Castillon2025}, Specimen 3 is omitted from this study. All simulations are performed under plane strain conditions with a specimen thickness of $B = 0.08W$.

The boundary conditions, illustrated in Figure~\ref{fig:compact_specimen_configuration}, involve constraining the lower half of the bottom loading hole in all directions and applying a distributed vertical force to the upper half of the top loading hole. Following the analysis in \cite{phase_field_Castillon2025}, a length scale parameter of $l = 0.0025W = 0.1$~mm is used with a mesh resolution of $l/h = 2.5$. The energy-controlled solver is employed with constants $c_1 = 1.0$~kN/mm and $c_2 = 1.0$ (dimensionless).

\begin{figure}[h!]
   \centering
   \subfigure[Specimen 1]{
      \includegraphics[width=0.30\linewidth]{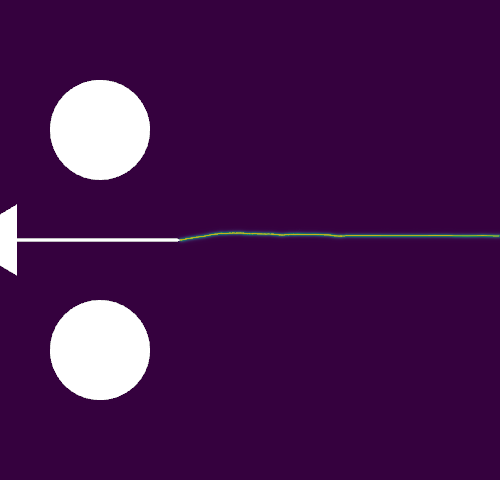}
      \label{fig:path_specimen2}
   }
   \hfill
   \subfigure[Specimen 2]{
      \includegraphics[width=0.30\linewidth]{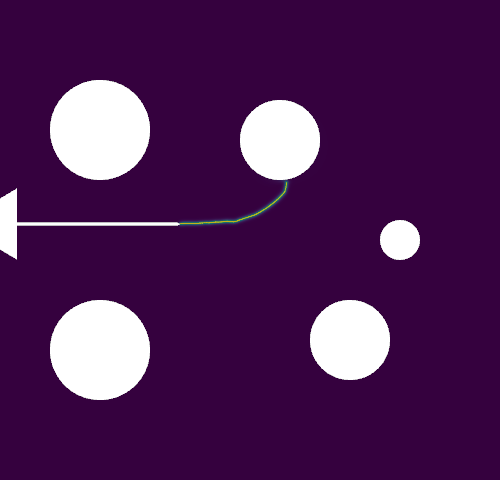}
      \label{fig:path_specimen3}
   }
   \hfill
   \subfigure[Specimen 4]{
      \includegraphics[width=0.30\linewidth]{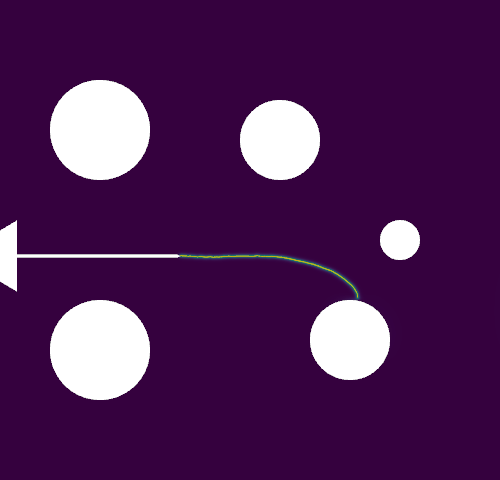}
      \label{fig:path_specimen4}
   }
      \caption{Final crack paths represented by the phase-field variable $\phi$ at the end of the simulation for: (a) Specimen 1 ($H=0.60W$), (b) Specimen 2 ($H=0.56W$), and (c) Specimen 4 ($H=0.64W$).}
   \label{fig:compact_specimen_with_holes_crack_path}
\end{figure}

Figure~\ref{fig:compact_specimen_with_holes_crack_path} shows the final phase-field profiles. For Specimen 1, a slight upward deviation at the beginning of the crack path is observed, which can be attributed to mesh effects. For Specimen 2, the crack propagates toward the upper hole, whereas for Specimen 4, where the initial crack position is lower with respect to the others, the crack path terminates at the bottom hole.

\begin{figure}[h!]
   \centering
   \subfigure[Evolution of correction factors.]{
      \includegraphics[width=0.45\linewidth]{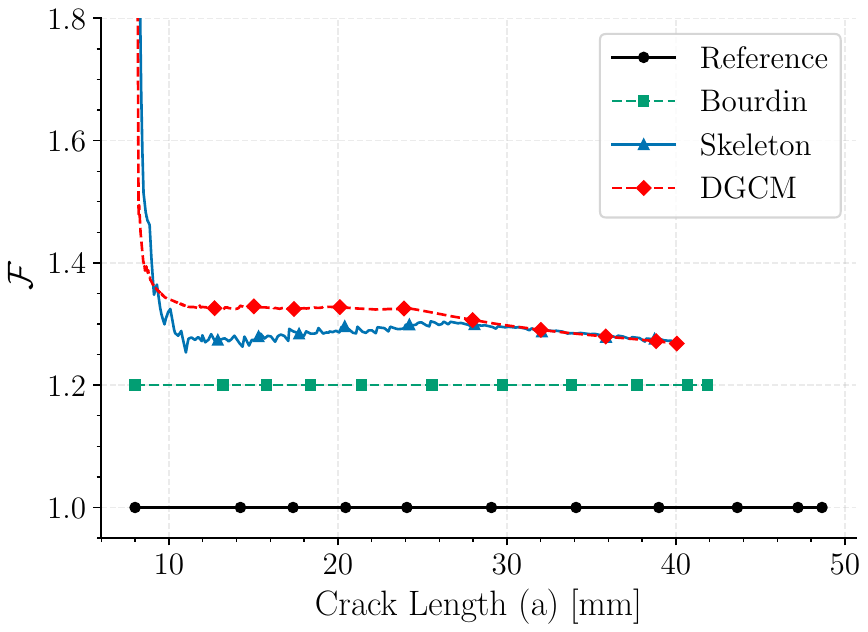}
      \label{fig:correction_crack_length_specimen_1}
   }
   \hfill
   \subfigure[Force-displacement response.]{
      \includegraphics[width=0.45\linewidth]{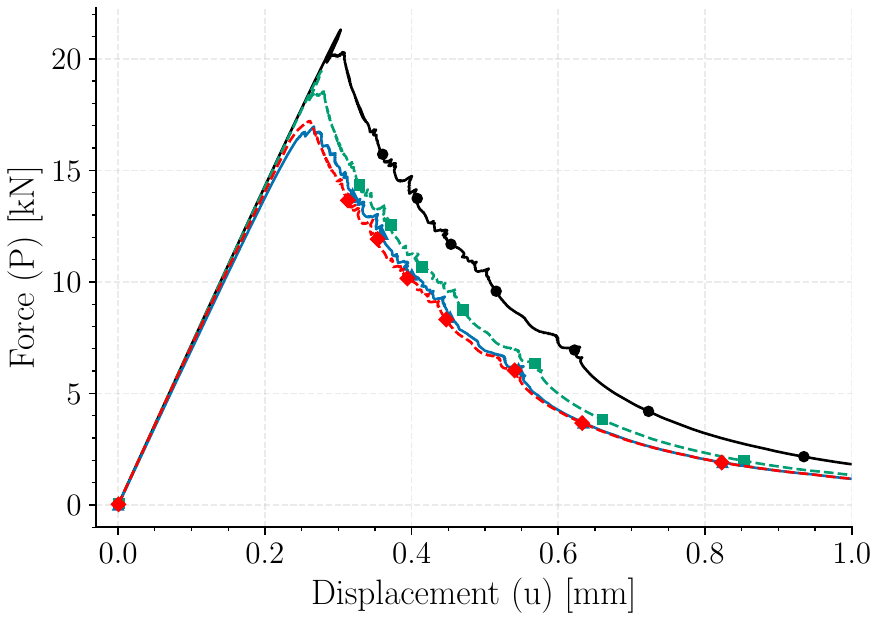}
      \label{fig:correction_force_diplacement_specimen_1}
   }
   \caption{Specimen 1: (a) Comparison of crack length correction factors from the Bourdin, Skeletonization, and DGCM methods against the uncorrected result. (b) Force-displacement curves corresponding to each correction method.}
   \label{fig:compact_specimen_1}
\end{figure}

For Specimen 1, Figure~\ref{fig:correction_crack_length_specimen_1} compares the correction factors obtained using the Bourdin, skeletonization, and the proposed DGCM methods, while Figure~\ref{fig:correction_force_diplacement_specimen_1} presents the corresponding force-displacement curves. The uncorrected simulation yields a final crack length of $48.65$~mm and a peak force of $21.31$~kN. Applying the corrections significantly modifies these values: the Bourdin method results in $41.88$~mm and $19.45$~kN, whereas the skeletonization and DGCM methods provide much closer estimates of $39.99$~mm ($16.96$~kN) and $40.06$~mm ($17.21$~kN), respectively.

\begin{figure}[h!]
   \centering
   \subfigure[Evolution of correction factors.]{
      \includegraphics[width=0.45\linewidth]{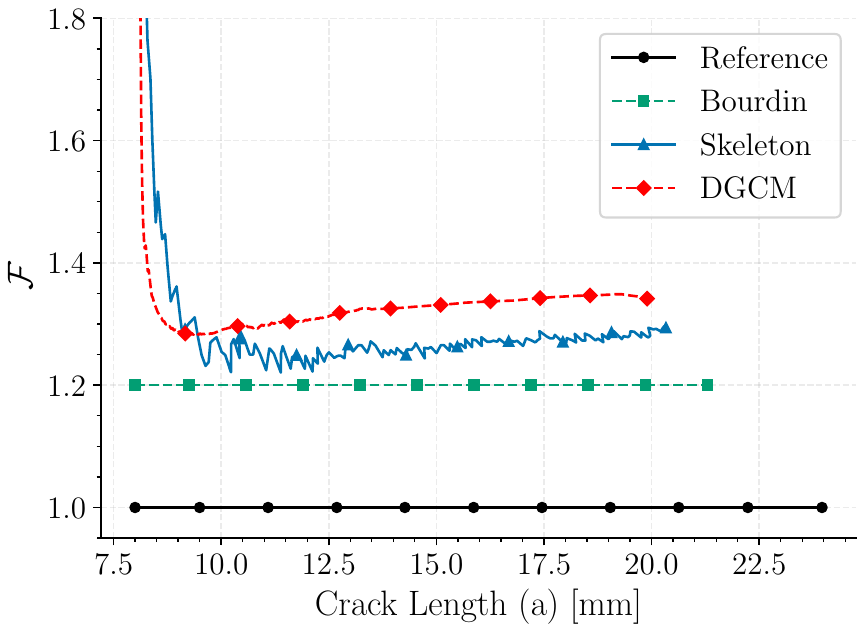}
      \label{fig:correction_crack_length_specimen_2}
   }
   \hfill
   \subfigure[Force-displacement response.]{
      \includegraphics[width=0.45\linewidth]{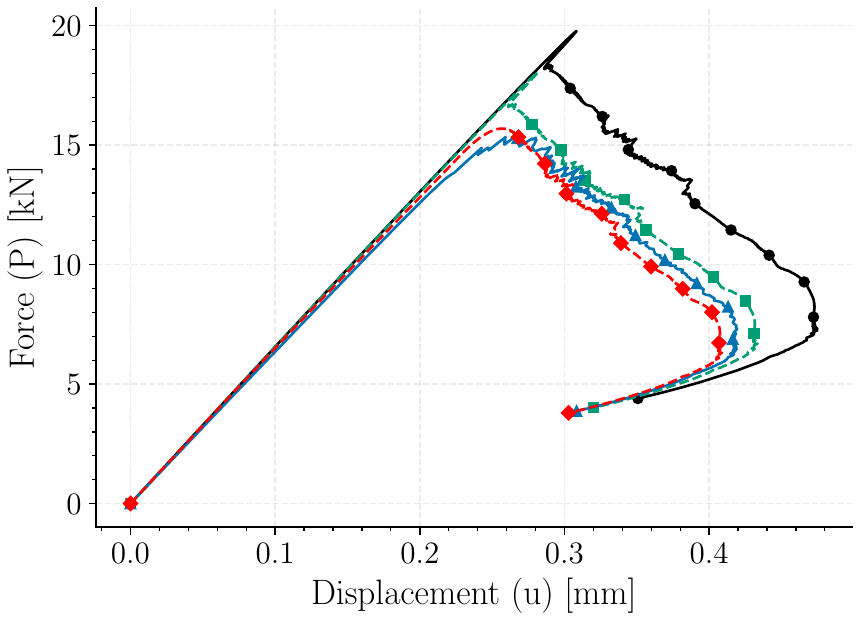}
      \label{fig:correction_force_diplacement_specimen_2}
   }
   \caption{Specimen 2: (a) Comparison of crack length correction factors from the Bourdin, Skeletonization, and DGCM methods against the uncorrected result. (b) Force-displacement curves corresponding to each correction method.}
   \label{fig:compact_specimen_2}
\end{figure}

Turning to Specimen 2, the correction factors and force-displacement responses are depicted in Figures~\ref{fig:correction_crack_length_specimen_2} and \ref{fig:correction_force_diplacement_specimen_2}, respectively. In the uncorrected simulation, the final crack length reaches $23.96$~mm with a peak force of $19.77$~kN. While the Bourdin correction adjusts these values to $21.30$~mm and $18.05$~kN, the skeletonization and DGCM methods yield even smaller and more consistent estimates: $20.35$~mm ($15.41$~kN) and $19.90$~mm ($15.70$~kN), respectively.

\begin{figure}[h!]
   \centering
   \subfigure[Evolution of correction factors.]{
      \includegraphics[width=0.45\linewidth]{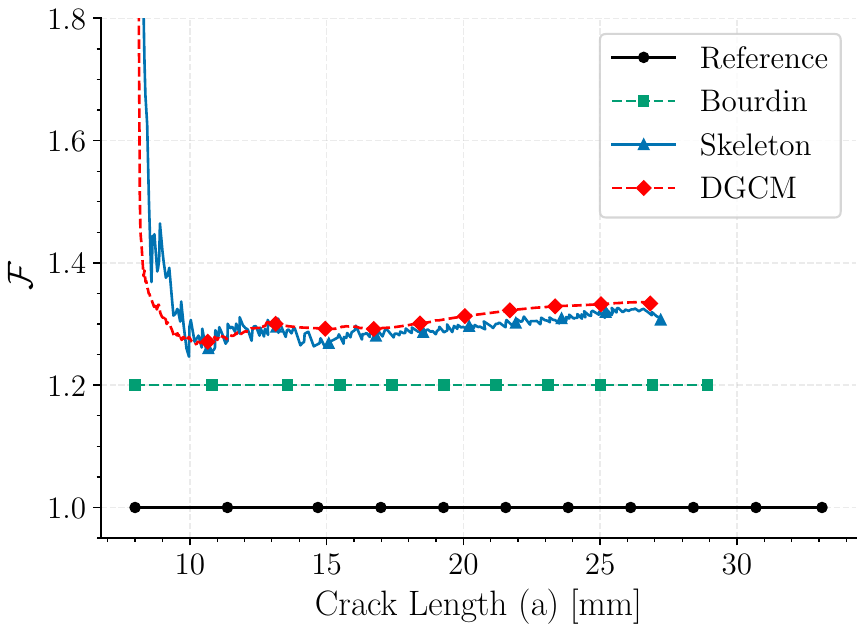}
      \label{fig:correction_crack_length_specimen_4}
   }
   \hfill
   \subfigure[Force-displacement response.]{
      \includegraphics[width=0.45\linewidth]{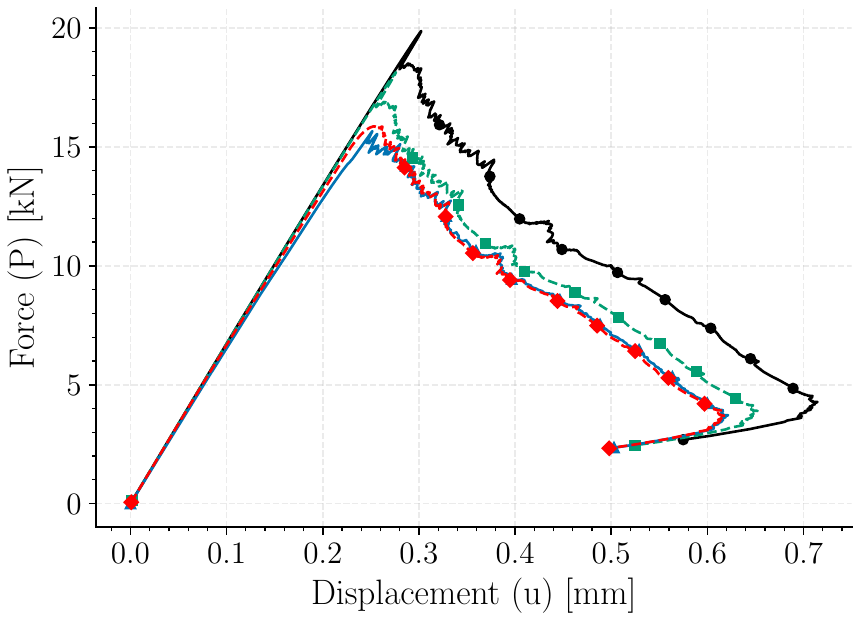}
      \label{fig:correction_force_diplacement_specimen_4}
   }
   \caption{Specimen 4: (a) Comparison of crack length correction factors from the Bourdin, Skeletonization, and DGCM methods against the uncorrected result. (b) Force-displacement curves corresponding to each correction method.}
   \label{fig:compact_specimen_4}
\end{figure}

Finally, for Specimen 4, the results are depicted in Figures~\ref{fig:correction_crack_length_specimen_4} and \ref{fig:correction_force_diplacement_specimen_4}. The uncorrected simulation predicts a final crack length of $33.11$~mm and a peak force of $19.87$~kN. The Bourdin correction adjusts these values to $28.93$~mm and $18.14$~kN. Consistent with the previous cases, the skeletonization and DGCM methods provide closer agreement, yielding corrected lengths of $27.22$~mm ($15.67$~kN) and $26.83$~mm ($15.87$~kN), respectively.

In all three cases, the uncorrected results exhibit the highest overestimation of the crack length. While the Bourdin correction reduces this overestimation, a significant discrepancy remains. In contrast, both the skeletonization and the proposed DGCM methods yield very similar and more accurate results.

A notable artifact is the presence of a peak force in the uncorrected force-displacement curves, which persists even when the Bourdin correction is applied. This occurs because the Bourdin factor acts as a constant scaling term dependent only on the mesh size and length scale. Conversely, the dynamic correction factors from the skeletonization and DGCM methods effectively eliminate this non-physical peak force. As demonstrated in the previous example, this regularization effect exhibits convergent behavior.






\section{Conclusions}
\label{sec:conclusions}

This paper introduces a novel method for correcting the overestimation of crack area in phase-field
fracture simulations. Accurate crack measurement is essential for the proper analysis of phase-field
fracture results, as this overestimation systematically affects the evaluation of key physical quantities such as forces, displacements, and energies.

The proposed correction factor is based on the following key observations:
\begin{itemize}
   \item In standard finite element implementations of phase-field fracture models, a numerical artifact known as strain localization frequently occurs. This effect causes the phase-field variable, $\phi$, to artificially saturate at a value of 1 across entire elements along the crack path. Consequently, the energy component dependent on the phase-field variable is significantly overestimated, leading to an inflated measurement of the crack area via the crack surface density functional.
   \item Crucially, the energy component dependent on the phase-field gradient is not affected in the same way. Within a saturated element, the gradient of the phase field is zero; therefore, this artifact does not introduce an overestimation in the gradient-dependent energy term.
   \item The energy contributions from the phase-field variable and its gradient are equal, provided there is sufficient space for the phase field to develop without interference from boundaries or other cracks, a condition achieved when the length scale is sufficiently small. This is also illustrated in the one-dimensional case, where for an $a/l$ ratio of 5, the error in this equipartition is less than $0.1\%$. Consequently, the physical crack area can be accurately approximated as twice the gradient-dependent energy. This result forms the basis of the proposed correction factor.
\end{itemize}

This correction method enables accurate fracture mechanics analysis without a significant increase in computational cost, as it relies on energy quantities already computed during the simulation. It is directly applicable to three-dimensional simulations, avoiding the need for complex skeletonization algorithms. Furthermore, its definition is independent of the finite element type (e.g., triangles, quadrilaterals, tetrahedra, or hexahedra), although the analyses in this work were conducted using quadrilateral and tetrahedral elements.

It is important to note that both the Bourdin and DGCM corrections are valid only when the length scale parameter is sufficiently small to satisfy the condition of energy equipartition. This requires the crack length to be large enough relative to the length scale to allow the phase-field to develop without interference from boundaries or other cracks. In the context of the Bourdin correction, this is equivalent to ensuring sufficient space for regularization to approach the theoretical infinite-domain solution.

Regarding the peak force artifact observed in uncorrected force-displacement curves, the proposed DGCM effectively mitigates this issue. By dynamically adjusting the correction factor based on the evolving energy distribution, the DGCM eliminates non-physical peaks. It is worth noting that the peak force predicted by this method is generally lower than that of other methods, yielding a more conservative estimate of the maximum load. Convergence studies demonstrate that as the length scale parameter is reduced, the peak force obtained with the proposed correction converges to a stable value. Additionally, the proposed method exhibits less sensitivity to mesh resolution for a constant length scale compared to the Bourdin correction, as the latter depends explicitly on the element size.

Another critical condition is that the mesh resolution must be sufficient to accurately capture the phase-field profile. As established by Miehe et al.~\cite{phase_field_Miehe_lh_relation}, a ratio of $l/h \ge 2$ is generally recommended for linear elements. Our analysis confirms that for this resolution, the error in the crack surface density functional is approximately $1\%$. While this error analysis is based on the one-dimensional case, the strain localization effect—which induces a constant phase-field value of $\phi=1$ across an entire element—implies that the discretization error is directly related to the symmetric problem analyzed previously. Therefore, the resolution criteria discussed ensure a correct approximation, though extensions to two or three dimensions may introduce additional geometric considerations.

It is also important to note that all simulations in this work were performed using an energy-controlled solver. This approach enables the capture of the full equilibrium path during crack propagation. Furthermore, this algorithm enforces irreversibility globally, ensuring that the sum of the fracture energy components does not decrease. This condition allows for the analysis of strain localization without the constraints of local irreversibility enforcement. While this study traces the equilibrium path, it is worth noting that in standard displacement-controlled schemes, the critical load would correspond to the peak force overshoot discussed herein.

While this work focuses on the widely used AT2 crack surface representation, the methodology can be adapted for other regularized models, such as the AT1 model, as detailed in Appendix~\ref{sec:appendix_general_phase_field}. For models with finite support like AT1, the application of the correction factor is valid provided the phase-field support is fully developed—that is, when the crack length exceeds the support width of the phase-field. Under these conditions, energy equipartition holds, and the correction factor can be applied.

\section{Code availability}
\label{sec:code_availability}
To ensure transparency and reproducibility, the complete set of source code, simulation scripts, and mesh files used in this work is publicly available. The repository is hosted on GitHub~\cite{code_availability} and permanently archived on Zenodo~\cite{code_availability_zenodo}. The numerical simulations were performed using `PhaseFieldX`~\cite{code_phasefieldx} (v0.2.0~\cite{code_phasefieldx_version}), an open-source library built upon the FEniCSx finite element framework~\cite{code_BarattaEtal2023}. All results are fully reproducible with the provided materials.

\section{Acknowledgements}
This work was partially supported by the Spanish Ministry of Science and Innovation through the FPI
grant PRE2020-092051 (MCIN/AEI/10.13039/501100011033). J.S. and I.R. were partially supported by the
Spanish Ministry of Science and Innovation through project IRIDISCENTE (ref. PLEC2023-010190).

\section{Declaration of generative ai and ai-assisted technologies in the writing process}
During the preparation of this work, the author(s) used Gemini 2.5 Pro to improve the readability and language of the manuscript. After using this tool, the author(s) reviewed and edited the content as needed and take full responsibility for the content of the published article.

\appendix
\renewcommand{\theequation}{A.\arabic{equation}}
\setcounter{equation}{0}

\section{Extension to general phase-field formulations}
\label{sec:appendix_general_phase_field}
Although the derivation presented in this work focuses on the AT2 model (Eq.~\eqref{eq:general_phase_field}), the proposed correction method is applicable to a broader class of phase-field formulations. These models typically share a common structure comprising a phase-field dependent term and a gradient dependent term. They are constructed such that, when the phase-field profile is fully developed and free from boundary effects, the total crack surface energy integrates to 1, and the energy contributions from the phase-field and gradient terms satisfy the equipartition result.

A general form for the crack surface density functional can be expressed as:
\begin{equation}
   \Gamma[\phi] = \frac{1}{c_0} \int_\Omega \left( \frac{\alpha(\phi)}{l} + l |\nabla \phi|^2 \right) d\Omega,
\end{equation}
where $c_0$ is a normalization constant and $\alpha(\phi)$ is a geometric function specific to the model. For the AT2 model used in the main text, $\alpha(\phi) = \phi^2$ and $c_0 = 2$. For the AT1 model, which features a finite elastic domain, $\alpha(\phi) = \phi$ and $c_0 = 8/3$.

Following the one-dimensional analysis, the total energy in the absence of numerical artifacts is given by:
\begin{equation}
    \Gamma_{\text{t}}[\phi] = \frac{1}{c_0} \int_{-a}^{+a}  \left( \frac{\alpha(\phi)}{l} + l (\phi')^2 \right) dx.
\end{equation}

When strain localization occurs, the phase-field variable artificially saturates to $\phi=1$ across an element of size $h$, causing the gradient to vanish locally ($\phi'=0$). The computed energy, $\Gamma_{\text{sl}}$, consequently includes an erroneous contribution from this saturated region:
\begin{equation}
    \Gamma_{\text{sl}}[\phi] = \Gamma_{\text{t}}[\phi] + \frac{1}{c_0} \int_{0}^{h} \frac{\alpha(1)}{l}  \, dx = \Gamma_{\text{t}}[\phi] + \frac{h}{c_0 l}.
\end{equation}
Assuming $\Gamma_{\text{t}} \approx 1$, this leads to a generalized Bourdin-like correction factor of $(1 + \frac{h}{c_0 l})$. This approximation holds when the phase-field has sufficient space to develop without boundary effects. For infinite support models like AT2, this requires $l \to 0$. However, for models with finite support like AT1, this condition is satisfied exactly provided the crack length exceeds the support width.

By relying on the energy equipartition property, a more robust correction can be formulated. Since the gradient term remains unaffected by the saturation artifact, the crack area can be approximated as twice the gradient energy:
\begin{equation}
    \Gamma \approx 2 \Gamma_{\nabla \phi} = \frac{2}{c_0} \int_\Omega l |\nabla \phi|^2 d\Omega.
\end{equation}

Consequently, the generalized Double Gradient Correction Method (DGCM) factor for any such functional is defined as:
\begin{equation}
\mathcal{F}_\mathrm{DGCM} = \frac{1}{2} \left[1 + \frac{1}{l^2}\frac{\int_\Omega \alpha(\phi) \,\mathrm{d}\Omega}{\int_\Omega |\nabla \phi|^2 \,\mathrm{d}\Omega}\right].
\end{equation}

\printbibliography 

@article{code_phasefieldx,
  author = {Castillón, M.}, 
  title = {PhaseFieldX: an open-source framework for advanced phase-field simulations}, 
  journal = {Journal of Open Source Software},
  volume = {10}, 
  number = {108}, 
  pages = {7307}, 
  year = {2025}, 
  doi = {https://doi.org/10.21105/joss.07307}
}

@software{code_phasefieldx_version,
  author       = {Castillón, M.},
  title        = {PhaseFieldX: An Open-Source Framework for Advanced Phase-Field Simulations},
  month        = oct,
  year         = 2025,
  publisher    = {Zenodo},
  version      = {v0.2.0},
  doi          = {10.5281/zenodo.17454862},
  url          = {https://doi.org/10.5281/zenodo.17454862},
  swhid        ={swh:1:dir:b32f8a9266d2a8e6b5b712612838db2f7de265eb;origin=https://doi.org/10.5281/zenodo.14972115;visit=swh:1:snp:7fa76041ea41e0ef1cd6e31f8b9fa9de65068c9f;anchor=swh:1:rel:802d99bef22fed3cc118d1cec98d779051908a00;path=CastillonMiguel-phasefieldx-2841493}
}

@misc{code_availability,
  author       = {Castillón, M.},
  title        = {{Github Repository to be added upon acceptance}},
}

@misc{code_availability_zenodo,
  author       = {Castillón, M.},
  title        = {{Zenodo DOI to be added upon acceptance}},
}

@misc{code_BarattaEtal2023,
  author    = {Baratta, I.A. and Dean, J.P. and Dokken, J.S. and Habera, M. and Hale, J.S. and Richardson, C.N. and Rognes, M.E. and Scroggs, M.W. and Sime, N. and Wells, G.N.},
  title     = {DOLFINx: the next generation FEniCS problem solving environment},
  doi       = {https://doi.org/10.5281/zenodo.10447666},
  year      = {2023}
}

@article{phase_field_Castillon2025,
title = {A phase-field approach to fatigue analysis: Bridging theory and simulation},
journal = {International Journal of Fatigue},
volume = {205},
pages = {109397},
year = {2026},
issn = {0142-1123},
doi = {https://doi.org/10.1016/j.ijfatigue.2025.109397},
url = {https://www.sciencedirect.com/science/article/pii/S0142112325005948},
author    = {Castillón, M. and Romero, I. and Segurado, J.},
}

@article{phase_field_Miehe2010,
  author    = {Miehe, C. and Hofacker, M. and Welschinger, F.},
  title     = {A phase field model for rate-independent crack propagation: robust algorithmic implementation based on operator splits},
  journal   = {Computer Methods in Applied Mechanics and Engineering},
  volume    = {199},
  number    = {45-48},
  pages     = {2765--2778},
  year      = {2010},
  doi       = {https://doi.org/10.1016/j.cma.2010.04.011}
}

@article{phase_field_Miehe_lh_relation,
author = {Miehe, C. and Welschinger, F. and Hofacker, M.},
title = {Thermodynamically consistent phase-field models of fracture: Variational principles and multi-field FE implementations},
journal = {International Journal for Numerical Methods in Engineering},
volume = {83},
number = {10},
pages = {1273-1311},
doi = {https://doi.org/10.1002/nme.2861},
url = {https://onlinelibrary.wiley.com/doi/abs/10.1002/nme.2861},
eprint = {https://onlinelibrary.wiley.com/doi/pdf/10.1002/nme.2861},
year = {2010}
}

@article{phase_field_FrancfortMarigo1998,
  author    = {Francfort, G.A. and Marigo, J.-J.},
  title     = {Revisiting brittle fracture as an energy minimization problem},
  journal   = {Journal of the Mechanics and Physics of Solids},
  volume    = {46},
  number    = {8},
  pages     = {1319--1342},
  year      = {1998},
  doi       = {https://doi.org/10.1016/S0022-5096(98)00034-9}
}

@article{phase_field_Bourdin2008,
  author    = {Bourdin, B. and Francfort, G.A. and Marigo, J.-J.},
  title     = {The Variational Approach to Fracture},
  journal   = {Journal of Elasticity},
  year      = {2008},
  volume    = {91},
  number    = {1},
  pages     = {5--148},
  doi       = {10.1007/s10659-007-9107-3},
  url       = {https://doi.org/10.1007/s10659-007-9107-3},
  issn      = {1573-2681}
}

@article{phase_field_Bourdin2000,
  author = {Bourdin, B. and Francfort, G.A. and Marigo, J.-J.},
  title = {Numerical experiments in revisited brittle fracture},
  journal = {Journal of the Mechanics and Physics of Solids},
  volume = {48},
  number = {4},
  pages = {797--826},
  year = {2000},
  doi = {https://doi.org/10.1016/S0022-5096(99)00028-9}
}

@article{phase_field_non_dimensional,
title = {Phase-field modeling of fracture with physics-informed deep learning},
journal = {Computer Methods in Applied Mechanics and Engineering},
volume = {429},
pages = {117104},
year = {2024},
issn = {0045-7825},
doi = {https://doi.org/10.1016/j.cma.2024.117104},
url = {https://www.sciencedirect.com/science/article/pii/S0045782524003608},
author = {M. Manav and R. Molinaro and S. Mishra and L. {De Lorenzis}},
}

@article{phase_field_snap_pedro,
  author = {Aranda, P. and Segurado, J.},
  title = {A crack-length control technique for phase-field fracture in {{FFT}} homogenization},
  journal = {International Journal for Numerical Methods in Engineering},
  volume = {126},
  number = {2},
  pages = {e7664},
  doi = {https://doi.org/10.1002/nme.7664},
  url = {https://onlinelibrary.wiley.com/doi/abs/10.1002/nme.7664},
  eprint = {https://onlinelibrary.wiley.com/doi/pdf/10.1002/nme.7664},
  year = {2025}
}

@article{phase_field_snap_Ritukesh,
  author = {Bharali, R. and Goswami, S. and Anitescu, C. and Rabczuk, T.},
  title = {A robust monolithic solver for phase-field fracture integrated with fracture energy based arc-length method and under-relaxation},
  journal = {Computer Methods in Applied Mechanics and Engineering},
  volume = {394},
  pages = {114927},
  year = {2022},
  doi = {https://doi.org/10.1016/j.cma.2022.114927}
}

@article{phase_field_snap_Zambrano,
  author    = {Zambrano, J. and Toro, S. and Sánchez, P.J. and Duda, F.P. and Méndez, C.G. and Huespe, A.E.},
  title     = {An arc-length control technique for solving quasi-static fracture problems with phase field models and a staggered scheme},
  journal   = {Computational Mechanics},
  year      = {2024},
  volume    = {73},
  number    = {4},
  pages     = {751--772},
  doi       = {https://doi.org/10.1007/s00466-023-02388-7}
}

@article{phase_field_skeleton,
  author = {Greco, L. and Patton, A. and Negri, M. and Marengo, A. and Perego, U. and Reali, A.},
  title = {Higher order phase-field modeling of brittle fracture via isogeometric analysis},
  journal = {Engineering with Computers},
  year = {2024},
  volume = {40},
  number = {6},
  pages = {3541--3560},
  doi = {https://doi.org/10.1007/s00366-024-01949-5}
}

@article{phase_field_effective_Gc_factor_2,
  title = {Thermodynamically consistent linear-gradient damage model in Abaqus},
  journal = {Engineering Fracture Mechanics},
  volume = {266},
  pages = {108390},
  year = {2022},
  issn = {0013-7944},
  doi = {https://doi.org/10.1016/j.engfracmech.2022.108390},
  url = {https://www.sciencedirect.com/science/article/pii/S001379442200145X},
  author = {Molnár, G. and Doitrand, A. and Jaccon, A. and Prabel, B. and Gravouil, A.},
}

@phdthesis{example_Wagner2018_phd_thesis,
  author = {Wagner, D.},
  title = {A finite element-based adaptive energy response function method for curvilinear progressive fracture},
  school = {The University of Texas at San Antonio},
  year = {2018},
  address = {United States -- Texas},
  pages = {181}
}

@article{example_Wagner2019_article,
  title = {A finite element-based adaptive energy response function method for 2D curvilinear progressive fracture},
  journal = {International Journal of Fatigue},
  volume = {127},
  pages = {229-245},
  year = {2019},
  issn = {0142-1123},
  doi = {https://doi.org/10.1016/j.ijfatigue.2019.05.036},
  url = {https://www.sciencedirect.com/science/article/pii/S0142112319302300},
  author = {Wagner, D. and Garcia, M.J. and Montoya, A. and Millwater, H.},
}

@Book{hughes1987vn,
  author       = {Hughes, T. J. R.},
  title        = {The finite element method},
  publisher    = {Prentice-Hall Inc.},
  year         = 1987,
  address      = {Englewood Cliffs, New Jersey},
  isbn         = {0-13-317025-X},
}

@Book{ern2004wx,
  author       = {Ern, A. and Guermond, J. L.},
  title        = {{Theory and practice of finite elements}},
  year         = 2004,
  publisher    = {Springer},
}

\end{document}